\documentclass[aps,prl,twocolumn,superscriptaddress]{revtex4-2}
\usepackage[T1]{fontenc} 
\usepackage{float}
\usepackage{graphicx}
\usepackage{dcolumn}
\usepackage{bm}
\usepackage{color}
\usepackage{amsmath}
\usepackage{todonotes}
\usepackage{enumitem}
\usepackage{amssymb}
\usepackage{xcolor}
\usepackage{braket}
\usepackage[colorlinks = true,
            linkcolor = blue,
            urlcolor  = red,
            citecolor = blue,
            anchorcolor = blue]{hyperref}
\usepackage{mathdots}

\begin{document}
\title{Evidence for chiral superconductivity on a silicon surface}

\author{F. Ming}
\affiliation{State Key Laboratory of Optoelectronic Materials and Technologies, School of Electronics and Information Technology and Guangdong Province Key Laboratory of Display Material, Sun Yat-sen University, Guangzhou 510275, China\looseness=-1}

\author{X. Wu}
\affiliation{Department of Physics, Southern University of Science and Technology, Shenzhen, Guangdong 518055, China\looseness=-1}
\affiliation{School of Physical Sciences, Great Bay University, Dongguan, Guangdong 523000, China\looseness=-1}

\author{C. Chen}
\affiliation{Department of Physics, Southern University of Science and Technology, Shenzhen, Guangdong 518055, China\looseness=-1}

\author{K. D. Wang}
\affiliation{Department of Physics, Southern University of Science and Technology, Shenzhen, Guangdong 518055, China\looseness=-1}

\author{P. Mai}
\affiliation{Computational Sciences and Engineering Division, Oak Ridge National Laboratory, Oak Ridge, TN, 37831-6494, USA\looseness=-1}

\author{T. A. Maier}
\affiliation{Computational Sciences and Engineering Division, Oak Ridge National Laboratory, Oak Ridge, TN, 37831-6494, USA\looseness=-1}

\author{J. Strockoz}
\affiliation{Department of Physics, Drexel University, Philadelphia, PA 19104, USA}
\affiliation{Department of Materials Science and Engineering, Drexel University, Philadelphia, PA 19104, USA\looseness=-1}

\author{J.~W.~F.~Venderbos}
\affiliation{Department of Physics, Drexel University, Philadelphia, PA 19104, USA}
\affiliation{Department of Materials Science and Engineering, Drexel University, Philadelphia, PA 19104, USA\looseness=-1}

\author{C.~Gonzalez}
\affiliation{Departamento de Física de Materiales, Universidad Complutense de Madrid, 28040 Madrid, Spain\looseness=-1}
\affiliation{Instituto de Magnetismo Aplicado UCM-ADIF, Vía de Servicio A-6, 900, E-28232 Las Rozas de Madrid, Spain\looseness=-1}

\author{J.~Ortega}
\affiliation{Departamento de F{\'i}sica Te{\'o}rica de la Materia Condensada and Condensed Matter Physics Center (IFIMAC), Facultad de Ciencias, Universidad Aut{\'o}noma de Madrid, 28049 Madrid, Spain\looseness=-1}

\author{S.~Johnston}
\affiliation{Department of Physics and Astronomy, The University of Tennessee, Knoxville, TN 37996, USA\looseness=-1}
\affiliation{Institute of Advanced Materials and Manufacturing, The University of Tennessee, Knoxville, TN 37996, USA\looseness=-1} 

\author{H.~H.~Weitering}
\affiliation{Department of Physics and Astronomy, The University of Tennessee, Knoxville, TN 37996, USA\looseness=-1}
\affiliation{Institute of Advanced Materials and Manufacturing, The University of Tennessee, Knoxville, TN 37996, USA\looseness=-1} 

\begin{abstract}
Sn adatoms on a Si(111) substrate with 1/3 monolayer coverage form a two-dimensional triangular adatom lattice with one unpaired electron per site and an antiferromagnetic Mott insulating state. The Sn layers can be modulation hole-doped and metallized using heavily-doped $p$-type Si(111) substrates, and become superconducting at low temperatures. While the pairing symmetry of the superconducting state is currently unknown, the combination of repulsive interactions and frustration inherent to the triangular adatom lattice opens up the possibility for a chiral order parameter. Here, we study the superconducting state of Sn/Si(111) using scanning tunneling microscopy/spectroscopy and quasi-particle interference imaging. We find evidence for a doping-dependent $T_c$ with a fully gapped order parameter, the presence of time-reversal symmetry breaking, and a strong enhancement of the zero-bias conductance near the edges of the superconducting domains. While each individual piece of evidence could have a more mundane interpretation, our combined results suggest the tantalizing possibility that Sn/Si(111) is an unconventional chiral $d$-wave superconductor. 
\end{abstract}

\maketitle

Superconductivity -- dissipationless electrical conductivity in conjunction with perfect diamagnetism -- is a profound manifestation of a macroscopic quantum phenomenon. Microscopically, supercurrents are carried by Cooper pairs whose pair wave functions become phase locked as they condense, like bosons, into a coherent macroscopic quantum state~\cite{BCS}. In conventional superconductors, electron pairing is mediated by virtual phonon exchange. In this case, the relatively slow motion of the ions provides a time-retarded effective attractive interaction that allows the electrons to overcome their mutual repulsion resulting in Cooper pairs with $s$-wave symmetry, where the composite spin and orbital angular momenta of the electrons are zero. Higher angular momentum states are typically driven by repulsive interactions~\cite{KohnPRL1965, ScalapinoRMP2012} as is the case for e.g. high-$T_c$ cuprate superconductors~\cite{ScalapinoRMP2012, TsueiRMP2000}. Here, electron repulsion is minimized by imposing a nodal structure with corresponding sign change in the superconducting wave function. More recent emphasis on topological materials systems have raised expectations for the discovery of novel multi-component order parameters that are topologically distinct from those of ordinary Cooper pair condensates~\cite{ReadPRB2000, JoyntRMP2002, NandkishoreNatPhys2012, Kallin2012, Black-SchafferPRL2012, KieselPRL2013, Black-SchafferJPCM2014, KallinRPP2016, Mackenzie2017, Pustogow2019, JiaoNature2020}.  Besides the microscopic nature of the pairing interactions, the physics of these systems is dictated by broken symmetries such as crystal, spin rotation, and time-reversal symmetries, though experimental validation of  intrinsically topological order parameters remains scant.

\begin{figure*}[t]
\centering
\includegraphics[width=\textwidth]{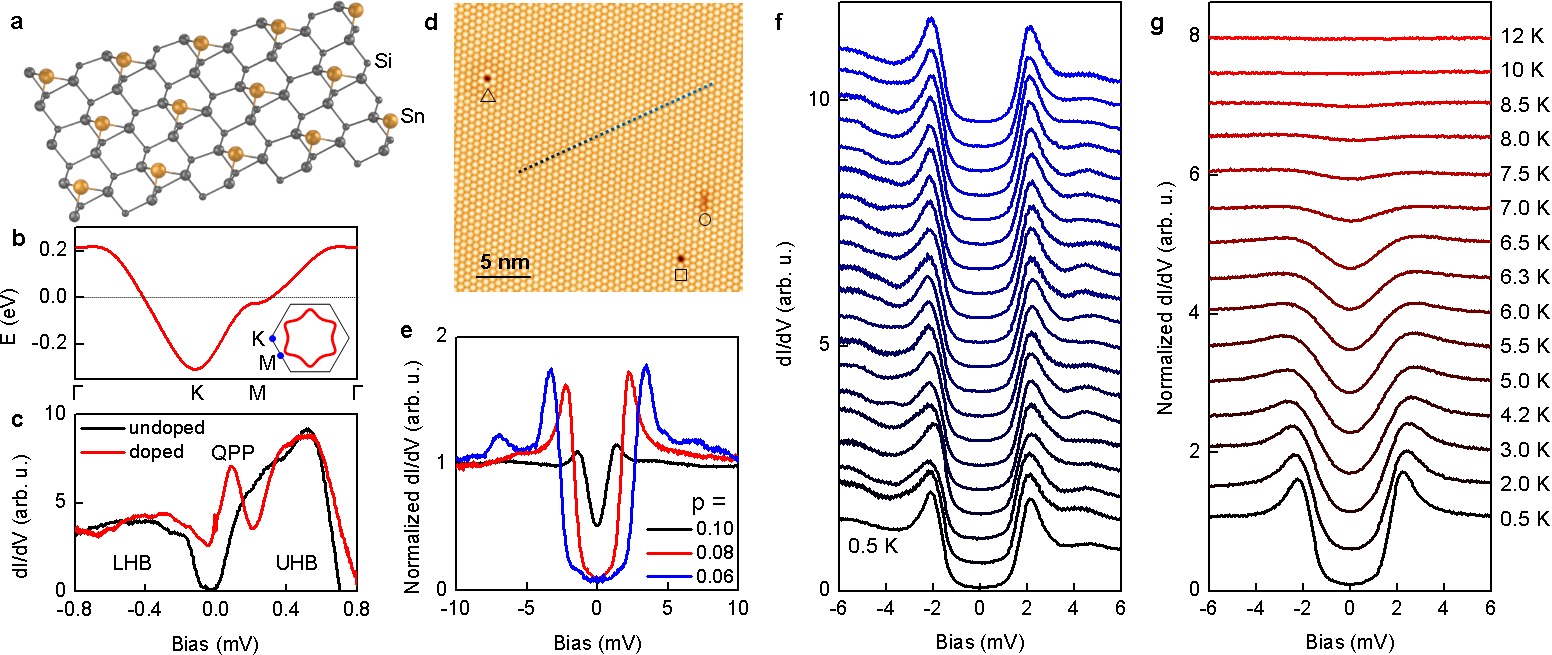}
\caption{{\bf Structure, topography and spectral properties of the superconducting $\mathbf{(\sqrt{3}\times\sqrt{3})}$-Sn surface on Si(111)}. {\bf a}, Structure model with the three outermost surface layers and one Sn adatom per $(\sqrt{3}\times\sqrt{3})$ unit cell. {\bf b}, Noninteracting dispersion of the dangling-bond surface state according to Ref.~\citenum{LiNatComm2013}. The inset shows the Fermi surface in the hexagonal surface Brillouin zone. The high symmetry points are indicated. The $\Gamma$ point is located at the center of the hexagon. {\bf c}, Comparison of differential conductance spectra of two Si(111)($\sqrt{3}\times\sqrt{3}$)-Sn surfaces. The spectrum for the undoped surface is from Ref.~\citenum{WuPRL2020}, acquired at 77 K, showing the upper and lower Hubbard bands (UHB/LHB) and a small gap around the Fermi level; the other spectrum is from a doped surface ($p=0.08$) obtained at 0.5 K, showing an extra quasiparticle peak (QPP) near the Fermi level, consistent with Ref.~\citenum{WuPRL2020}. {\bf d}, Topographic STM image ($V_s = -0.1$~V, $I_t = 0.1$~nA) showing a near perfect Sn adatom lattice, along with a substitutional Si defect, vacancy, and other defect, labelled with a triangle, square and a circle, respectively ($p=0.08$). {\bf e}, Normalized STS spectra for three different hole concentrations, revealing a clear doping dependence of the superconducting gap. {\bf f}, A set of raw $dI/dV$ spectra taken at equidistant locations along the dotted line shown in panel {\bf d}, starting on the left. ($p=0.08$). {\bf g}, Normalized $dI/dV$ spectra as a function of temperature ($p=0.08$). The spectra in panels {\bf f} and {\bf g} are offset vertically for clarity. 
}
\label{fig:1}
\end{figure*}

Superconductivity has recently been discovered in a system comprised of one-third monolayer of Sn deposited on degenerately doped $p$-type Si(111) substrates~\cite{WuPRL2020}. Its pairing symmetry, however,  remains undetermined. This system is of particular interest because the undoped Sn monolayer is an antiferromagnetic single-band Mott insulator~\cite{LiNatComm2013, MingPRL2017} that becomes superconducting upon hole doping, drawing interesting comparisons with the high-$T_c$ cuprates~\cite{ScalapinoRMP2012, LeeRMP2006} with $d$-wave order parameters. The Sn layer, however, has triangular lattice symmetry imposed by the Si(111) substrate. This geometry naturally allows for the existence of a chiral order parameter with topological edge states~\cite{NandkishoreNatPhys2012, KallinRPP2016, CaoPRB2018}, if repulsive interactions dominate the pairing. The appearance of such an exotic order parameter is expected to furthermore depend on the electron correlation strength, shape of the Fermi surface, and the doping level~\cite{NandkishoreNatPhys2012, CaoPRB2018, Wolf2022}. In particular, recent renormalization group calculations for the Sn/Si(111) system indicated a competition between chiral $d$- and $f$-wave and triplet $p$-wave instabilities, depending on the doping level and value of the nearest-neighbor Hubbard repulsion~\cite{Wolf2022}. At the same time, electron-phonon interactions, particularly to interfacial Si modes~\cite{ZahedifarPRB2019}, could drive a conventional $s$-wave pairing ~\cite{WuPRL2020}. 

Here we study the superconducting state of the Sn/Si(111) interface using scanning tunneling microscopy and spectroscopy (STM/STS) and quasiparticle interference (QPI) imaging. Our observations reveal a strong doping dependence of the superconducting $T_\mathrm{c}$, a fully gapped order parameter, the presence of time-reversal symmetry breaking, and a strong enhancement of the zero-bias conductance near the edges of the superconducting domains. While each of these observations may have a mundane explanation, we discuss why we believe that a chiral $d$-wave scenario offers the most consistent interpretation of the measurements and theoretical modeling. Final confirmation, however, awaits experimental validation concerning the topological nature of the edge-state conductance.

\begin{figure}[t]
\centering
\includegraphics[width=1\columnwidth]{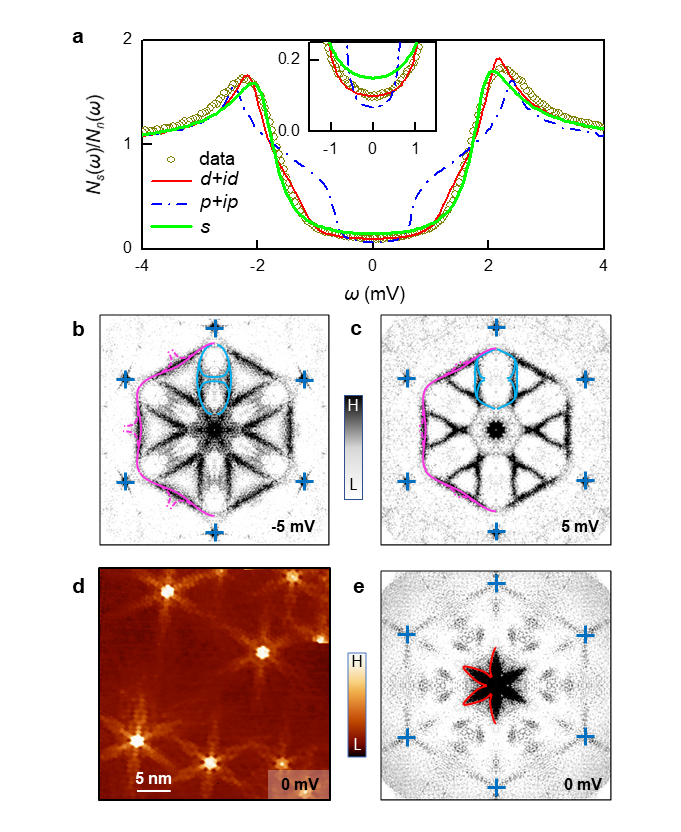}
\caption{{\bf Low-temperature differential conductance and QPI spectra of the ${\bf p}$ = 0.08 $\mathbf{(\sqrt{3}\times\sqrt{3})}$-Sn surface on Si(111)}. {\bf a}, Best fits of the low-energy normalized $dI/dV$ spectra at $T = 0.5$ K, for different pairing symmetries. The inset zooms in on the zero bias region. {\bf b}, {\bf c}, QPI images acquired at $V_s = \pm{5}$ mV, beyond the superconducting gap. {\bf d}, Real space conductance map $g({\bf r},V)$, obtained at zero bias. The bright six-leaf features are scattering features from surface defects. {\bf e}, Corresponding QPI spectrum $g({\bf q},V)$ obtained from the conductance map in panel {\bf d} via Fourier transformation. The six dark blue crosses in panels {\bf b}, {\bf c}, and {\bf e} indicate the Bragg peaks, while the colored contours highlight characteristic features in each QPI image. }
\label{fig:2}
\end{figure}

At 1/3 monolayer coverage, the Sn adatoms form a ($\sqrt{3}\times\sqrt{3}$) superlattice on the Si(111) surface with one half-filled dangling-bond orbital per site and a Sn-Sn distance of 6.65 \AA{}; see Fig.\ref{fig:1}{\bf{a}}. All other chemical bonds in the system are passivated. The non-interacting dangling-bond surface state has a bandwidth ${W} \approx{0.5}$ eV (Fig.~\ref{fig:1}{\bf b}), which is comparable to the on-site Hubbard interaction $U \approx  0.66$ eV of the dangling bond orbitals~\cite{LiNatComm2013}. As such, the system is a Mott insulator with an upper and a lower Hubbard band (UHB/LHB) straddling the Fermi level (Fig.~\ref{fig:1}{\bf c}).

Figure \ref{fig:1}{\bf{d}} shows an STM image of the triangular Sn adatom lattice. The Sn atoms are clearly resolved and well ordered. The dark point defects correspond to substitutional Si adatoms (most prevalent) and Sn adatom vacancies. Holes are introduced via modulation doping, using boron-doped Si substrates with different doping levels \cite{MingPRL2017,WuPRL2020}. (For a discussion on dopant segregation, see Supplementary Note 1 and Supplementary Figure 1). The hole concentration in the dangling-bond surface state is estimated from the spectral weight transfer in the tunneling spectra, associated with the introduction of holes and formation of a quasiparticle peak in the Mott gap (see Fig.~\ref{fig:1}{\bf b} and Extended Data Fig.~\ref{fig:ED1_QPPband}) \cite{MingPRL2017}. 

Fig.~\ref{fig:1}{\bf{e}} shows the normalized $dI/dV$ tunneling spectra for excess hole concentrations of $p=0.06$, $0.08$, and $0.10$, recorded at $T=0.5$~K. (The $p=0.10$ data were reported in Ref.~\citenum{WuPRL2020}.) These spectra are representative of the superconducting density of states (DOS). Here, we divided the raw tunneling spectra by the normal state $dI/dV$ spectrum obtained in a perpendicular $15$ Tesla magnetic field. This field is large enough to completely suppress the superconductivity for the $p=0.08$ and $0.10$ samples, which have upper critical field values of $H_{c2}(0.5)$ of 3 Tesla \cite{WuPRL2020} and 13 Tesla, respectively (see Extended Data Fig.~\ref{fig:superconducting_STS}). This procedure is a bit problematic for the $p=0.06$ sample, where the upper critical field exceeds the magnetic-field capability of our instrument (15 Tesla). 

The Sn adatom lattice in Figure \ref{fig:1}{\bf{a}} is highly ordered, while the boron dopants are located in the silicon bulk. The superconducting DOS is, therefore, spatially uniform away from localized point defects and the edges of the ($\sqrt{3}\times\sqrt{3}$) domains. Fig. \ref{fig:1}{\bf{f}} shows a series of spectra recorded at $\mathrm{T}= 0.5$~K along the line segment in Fig. \ref{fig:1}{\bf{d}}. This level of homogeneity distinguishes the Sn on Si(111) system from e.g. complex oxides, which exhibit considerable electronic inhomogeneity, often in conjunction with various competing orders~\cite{HowaldPNAS2003,Vershinin2004}. 

The normalized $dI/dV$ spectra of the $p=0.08$ sample are plotted as a function of temperature in Fig. \ref{fig:1}{\bf{g}}. The gap feature persists up to about $8$~K. Detailed Dynes fits~\cite{DynesPRL1978} of the spectra assuming $s$-wave and $d_{x^2-y^2}\pm\mathrm{i}d_{xy}$ order parameter symmetries, as well as zero bias conductance measurements as a function of temperature, consistently produce a $T_\mathrm{c}$ of about $7.6\pm{0.2}$~K with some evidence of superconducting fluctuations above $T_\mathrm{c}$ (see Ref.~\citenum{WuPRL2020}, Extended Data Fig.~\ref{fig:fitting_2delta}, and Fig. \ref{fig:2}{\bf a}). A similar procedure for the $p=0.10$ sample produces $T_\mathrm{c}$ $ = 4.7 \pm 0.3$~K \cite{WuPRL2020}, while the $T_\mathrm{c}$ of the $p=0.06$ sample was difficult to ascertain because the spectra cannot be properly normalized [$H_{c2}(0.5~\mathrm{K}) > 15$~T]. We conservatively estimate its $T_\mathrm{c}$ to be around $9$~K (see Extended Data Fig.~\ref{fig:superconducting_STS}). 

Fitting the $p=0.08$ $dI/dV$ spectra with an $s$-wave gap produces a reasonable fit but with notable discrepancies near zero-bias. Turning to potential chiral order parameters, we find that a chiral $d$-wave fit also agrees well with the data and even improves the fit at low energies. A chiral $p$-wave gap clearly fails to describe the spectra, particularly at low-energy (Fig. \ref{fig:2}{\bf{a}}). This failure occurs because any $p$-wave gap function must vanish at the $M$-point by symmetry. This point corresponds to the van Hove singularity and lays close to the Fermi surface~\cite{MingPRL2017, WuPRL2020}. It therefore affects the gap significantly, producing pronounced shoulders in the DOS that are not observed experimentally. Other parameter symmetries such as extended chiral $p$-wave and nematic $d$-wave symmetries (i.e., $d_{x^2-y^2}$ and $d_{xy}$) do not fit the spectra either, see Supplementary Note 2. (A multigap order parameter can also be ruled out since this is a single-band system.) Only $s$- and chiral $d$-wave symmetries produce good results and it is not possible to conclusively discriminate between the two based on fitting alone.

Important details about the Fermi surface and order parameter symmetry can be obtained from spectroscopic STM imaging~\cite{Petersen2002JESRP}. Here, one acquires a spatial map of the differential tunneling conductance $g(\textbf{r},V)=dI(\textbf{r},V)/dV$. Such $dI/dV$ maps typically reveal the presence of electronic standing waves as  quasi-particles are scattered elastically by defects on the surface. The power spectrum of the differential conductance map -- the QPI spectrum -- then identifies the dominant scattering processes contributing to the standing wave pattern. In itinerant systems, these typically correspond to scattering wavevectors connecting different {\bf k}-points on the constant energy contours (corresponding to the imaging bias) $\mathbf{q} = 2\mathbf{k} \pm \mathbf{G}$, where $\textbf{G}$ is a reciprocal lattice vector of the $(\sqrt{3}\times\sqrt{3})$ adatom lattice. 

\begin{figure}[t]
\centering
\includegraphics[width=0.8\columnwidth]{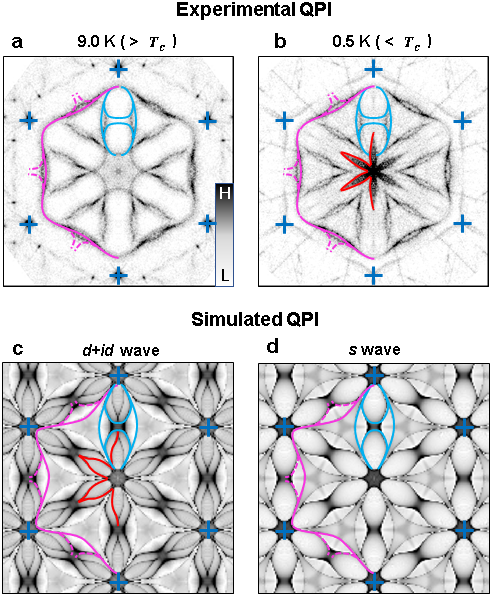}
\caption{{\bf Comparison of the measured QPI spectra with theory}. {\bf a}, {\bf b}, Experimental QPI images $g({\bf q},V)$ obtained at zero bias on the $p=0.10$ surface above (panel {\bf a}) and below $T_\mathrm{c}$ (panel {\bf b}). {\bf c}, {\bf d},
Simulated QPI images for a superconductor with a chiral $d$-wave (panel {\bf c}) and $s$-wave (panel {\bf d}) order parameter, assuming non-magnetic defects. In each image, the six dark blue crosses indicate the locations of the Bragg points, while the colored contours highlight characteristic features in each QPI spectrum. 
}
\label{fig:3}
\end{figure}

Figs. \ref{fig:2}\textbf{b}, {\bf c} show the $T = 0.5$~K QPI spectra taken at $\pm{5}$ meV bias ($p=0.08$). Both spectra reveal the warped hexagonal Fermi contour of the normal state ($\bf{G}=0$), highlighted in magenta, along with several scattering replica's ($\bf{G} \ne 0$) as indicated by the light blue dumbbell-shaped contours~\cite{MingPRL2017}.  These spectra agree very well with previous calculations for the spectra in the normal state, and are fully consistent with the band structure for the Sn surface state~\cite{MingPRL2017}. (The presence of the quasiparticle band and its dispersion is inconsistent with an interpretation in terms of impurity band physics; see Supplementary Note 3.) Real space differential conductance maps at zero bias (Fig.~\ref{fig:2}\textbf{d}), i.e. deep inside the superconducting gap, reveal the existence of very strong star-like scattering features centered at the various surface defects. The corresponding Fourier map (Fig.~\ref{fig:2}\textbf{e}) now reveals the presence of a flower-like feature centered at ${\bf q} = {\bf 0}$ and with six ``petals'' pointing towards the Bragg points of the ($\sqrt{3}\times\sqrt{3}$) lattice, as outlined by the red contour. Meanwhile, the Fermi contour seen at $\pm{5}$ meV is suppressed. This flower feature appears to be intimately related to the superconductivity; it only exists when the sample is in the superconducting state (Fig.~\ref{fig:3}{\bf a}, {\bf b}) and when the tunneling bias is within the superconducting gap (Fig.~\ref{fig:2}{\bf b}, {\bf c}, {\bf e}). Hence, they are unique features of the superconducting state (see Extended Data Fig.~\ref{fig:Experimental_QPI} for additional QPI results). 

To elucidate the origin of the flower-like features, we first simulated the QPI patterns for the $s$-wave and chiral $d$-wave order parameters using the T-matrix formalism and assuming ${\it nonmagnetic}$ scattering (see Methods). Fig.~\ref{fig:3} shows the experimental QPI spectra of the $p=0.10$ sample, along with the simulated spectra for the $d_{x^2-y^2}\pm\mathrm{i}d_{xy}$ and $s$-wave state.  The experimental pattern in Fig.~\ref{fig:3}{\bf b} is well reproduced in the calculations for the $d$-wave pairing channel (Fig.~\ref{fig:3}{\bf c}). (The theoretical features exhibit much more curvature because the calculations are based on the non-interacting band dispersion whereas the experimental band dispersion has correlation-driven band renormalizations \cite{MingPRL2017}.) Importantly, the flower is absent for the $s$-wave pairing channel, as shown in Fig.~\ref{fig:3}\textbf{d}. 

Our simulations reveal that the flower features only appear when time reversal symmetry is broken. Such would be the case for non-magnetic scattering in a chiral superconductor, as simulated above, but it could also be due to magnetic scattering in an $s$-wave superconductor (see Extended Data Fig.~\ref{fig:extended_QPI}). In particular, the star-like scattering features in the real-space QPI maps are very similar to those observed for magnetic point scatterers in $s$-wave systems, and have been attributed to a focusing effect of magnetic bound states or Yu–Shiba–Rusinov (YSR) states due to Fermi surface anisotropy \cite{Yu, Shiba, Rusinov, MenardNP2015, KimNC2020}. To discriminate between the $s$-wave and $d$-wave scenarios, it is essential to establish the nature of the defects on the surface.

The most prevalent scattering defect on the surface is the substitutional Si adatom (replacing a Sn adatom). It shows up as a dark void in filled-state STM images and as a depressed adatom in the empty state images (see Extended Data Fig.~\ref{fig:Defect_and_Simulation}). This observation indicates that the $sp_z$-like dangling bond orbital of the Si atom is empty, and because the adatom forms three covalent backbonds with the Si substrate, the Si adatom is expected to be non-magnetic. This is confirmed via first-principles Density Functional Theory (DFT) total-energy calculations (which show that the Si adatoms are placed 0.6~\AA{} below the Sn adatoms) and by STM image simulations (see Methods and Extended Data Fig.~\ref{fig:Defect_and_Simulation}). In addition, spin-polarized DFT calculations (see Methods) confirm the nonmagnetic nature of this defect.

\begin{figure*}[t]
\centering
\includegraphics[width=1\textwidth]{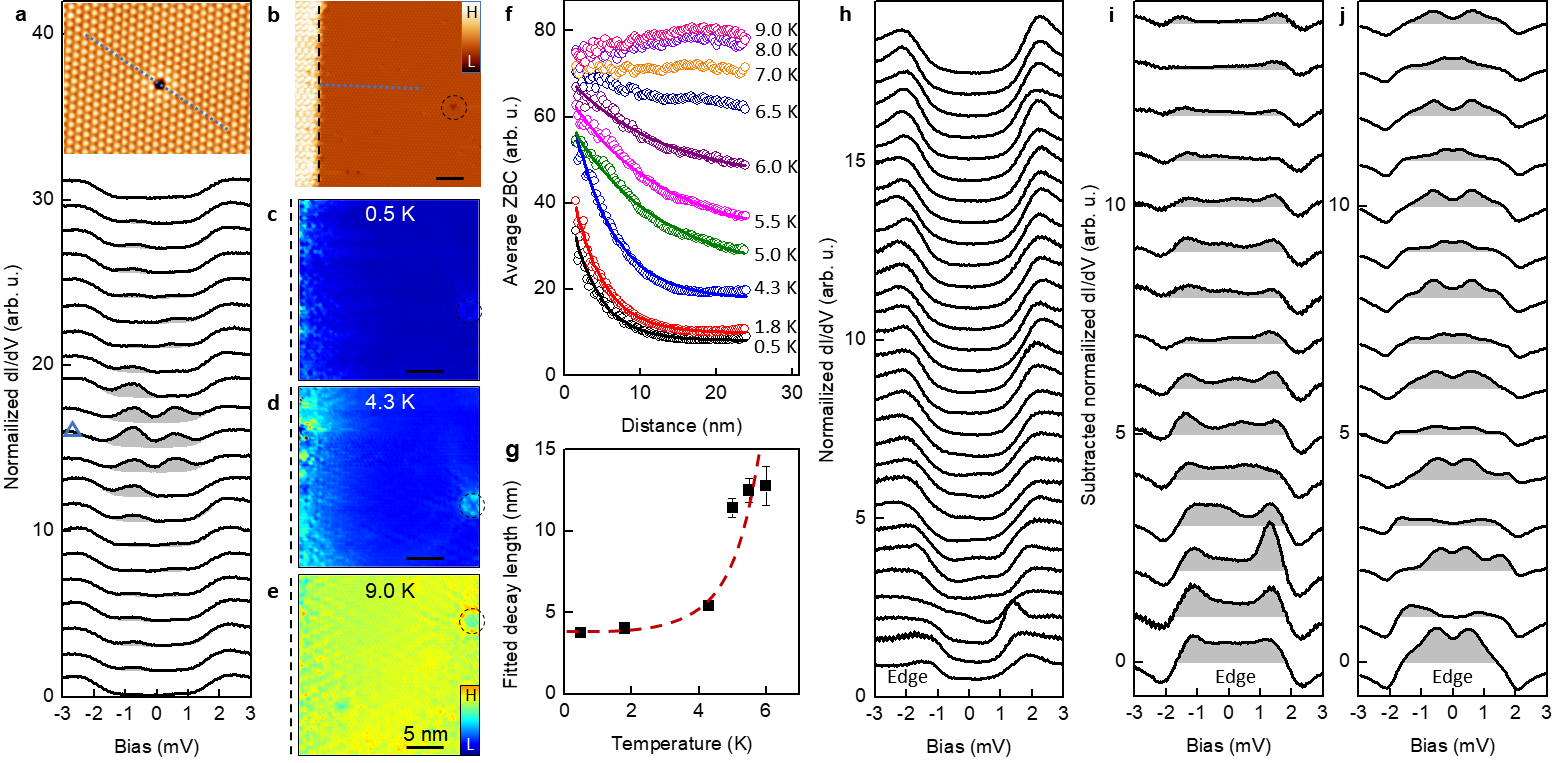}
\caption{{\bf Defect states and edge states on the $\mathbf{(\sqrt{3}\times\sqrt{3})}$-Sn surface ($\mathbf{p = 0.08}$)}. {\bf a}, STS point spectra (0.5~K) for each Sn atom along the blue-dotted line in the topographic image ($V_s$ = 0.5 V, $I_t$ = 1 nA) at the top with a substitutional Si defect in the middle. The bottom spectrum corresponds to the left end of the line. The spectrum recorded right on top of the defect is indicated by a triangle. Spectra near the defect site exhibit two gap states at $E_\mathrm{B}=\pm{0.6}$ meV. 
{\bf b}, A $(\sqrt{3}\times\sqrt{3})$-Sn superconducting domain ($V_s = 0.8$ V, $I_t = 0.1$ nA) next to a semiconducting Si(111)$(2\sqrt{3}\times 2\sqrt{3})$R30$^\circ$-Sn domain on the far left side of the image (bright strip) . {\bf c}-{\bf e}, Registry-aligned real-space conductance maps $g({\bf r},V = 0~\mathrm{mV})$ of the $(\sqrt{3}\times\sqrt{3})$-Sn domain measured at different temperatures. In panels {\bf b}-{\bf e}, the vertical dashed lines label the domain boundary; the dashed circles label the locations of the same defect.  {\bf f}, Averaged zero-bias conductance as a function of distance from the domain boundary. Each line is obtained from a conductance map $g({\bf r},V = 0~\mathrm{mV})$, recorded at the indicated temperature. The first six curves are fitted with an exponential decay. {\bf g}, Fitted decay lengths as a function of temperature. The dashed line is a guide to the eye. {\bf h}, STS spectra taken at 0.5~K along the dotted line in panel {\bf b}, starting at the domain boundary on the left. {\bf i}, The 15 bottommost spectra from panel {\bf h} after subtracting the $dI/dV$ spectrum recorded deep inside the $(\sqrt{3}\times\sqrt{3})$ domain. These spectra highlight the edge state contribution (indicated by shading) to the measured $dI/dV$ spectra. {\bf j}, Simulated DOS of a chiral $d+\mathrm{i}d$ superconductor, approaching the open edge of a cylinder (see Methods). As with panel {\bf i}, a spectrum from the center of the cylinder is subtracted to highlight the contribution from the edge state to the total DOS inside the superconducting gap. Spectra in panel {\bf h}, {\bf i} and {\bf j} are shifted vertically for clarity. }
\label{fig:4}
\end{figure*}

It is not possible to ascertain the nature of all native defects on the surface (see Extended Data Fig.~\ref{fig:VariousDefects}{\bf a}) and thus rule out any magnetic scattering contribution to the QPI pattern. We therefore created a new type of defect by depositing a tiny excess amount of (nonmagnetic) Sn  atoms at 120~K. STM images indicate that additional Sn adatoms are located at three-fold symmetric interstitial adatom sites, surrounded by three Sn adatoms of the 2D host lattice (see Extended Data Fig.~\ref{fig:Defect_and_Simulation}). The excess Sn atoms easily move under the STM tip at tunneling biases in excess of $\pm{0.8}$ V, indicating that they are weakly bound to the surface. The interstitial adatom location is validated by DFT total energy minimization and STM image simulations. In particular, notice the excellent agreement between the experimental and theoretical STM images, as shown in Extended Data Fig. ~\ref{fig:Defect_and_Simulation}. This level of agreement gives us confidence that the calculations capture the local electronic structure correctly. Importantly, spin-polarized DFT indicates that these defect centers are also nonmagnetic. The latter can be understood from the fact that the interstitial Sn atom and its three nearest neighbors have negligible contribution to the DOS at the Fermi level (Extended Data Fig.~\ref{fig:Defect_and_Simulation}), thus pre-empting a potential magnetic instability. 

Figure~\ref{fig:4}\textbf{a} plots the STS spectra across a substitutional Si defect, which reveals the existence of two in-gap states ($p=0.08$). These states appear in pairs located symmetrically about the Fermi level reminiscent of YSR bound states though the substitutional defect is nonmagnetic. All defects on the surface we have checked produce these YSR-like in gap states albeit at different energies, including the native vacancy defects, interstitial Si, excess Sn adatoms, as well as various other defects (see Extended Data Fig.~\ref{fig:VariousDefects}). This is to be expected in a chiral $d$-wave scenario because potential and magnetic defects are both pair breaking~\cite{WangPRB2004, MashkooriSSR2017}. In the $s$-wave scenario, on the other hand, one would have to assume that these defects are all magnetic~\cite{Anderson}, which seems unlikely (see Extended Data Fig.~\ref{fig:extended_QPI}). Interestingly, the nonmagnetic interstitial Sn adatoms produce the strongest star-like scattering features in real-space QPI maps and the most intense flower features in the corresponding power spectrum (see Extended Data Fig.~\ref{fig:VariousDefects}{\bf e-g}).  These enhanced scattering features near the excess Sn adatoms suggest that the $s$-wave scenario can be dismissed because time-reversal symmetry should not be broken in such a case. Alternatively, one might suggest that nonmagnetic impurities are present in a magnetically ordered background, where they behave as magnetic impurities. Such a scenario, however, would  require coexisting magnetism and superconductivity and point toward an unconventional order parameter (see also Supplementary Note 4). 

Theory predicts~\cite{WangPRB2004,MashkooriSSR2017} that the bound states of strong scatters will be located deeper inside the gap as compared to those of weaker scatterers. Indeed, we find that the substitutional Si defects produce in-gap states at $\pm{0.6}$~meV while interstitial Sn atoms produce states at $\pm{0.2}$~meV. Note that the existence of gap states alone for each and every scattering defect is an indication of unconventional (i.e., non-$s$-wave) superconductivity.

Our experimental measurements thus far point to a chiral $d$-wave superconducting state. To demonstrate that this pairing symmetry is consistent with the known electronic structure, we performed quantum Monte Carlo dynamical cluster approximation (DCA) method~\cite{MaierPRL2006, UrsCompPhysComm2020} calculations for the leading pairing instability of the ($\sqrt{3}\times\sqrt{3}$)-Sn system. Here, we consider a $3\times 3$ triangular lattice single-band Hubbard cluster embedded in a dynamical mean-field and adopt parameters previously used to describe doped~\cite{MingPRL2017} and undoped~\cite{LiNatComm2013} ($\sqrt{3}\times\sqrt{3}$)-Sn (see Methods). Since our calculations are limited by the Fermion sign problem, we focused on doping levels of $p = 0.05$, $0.10$, and $0.15$, where we are able to access temperatures as low as $T = 6.59$~meV $(\beta t_1 =  8$). 
For these parameters, the dominant superconducting instability indeed corresponds to degenerate $d_{x^2-y^2}$ and $d_{xy}$ order parameters at all doping levels (see Extended Data Fig.~\ref{fig:dca_eigenvectors}). 
These results are consistent not only with our observations, but also several prior studies for the triangular lattice Hubbard model~\cite{ChenPRB2013, CaoPRB2018} and a recent renormalization group study for Sn/Si(111) that found chiral $p$-, $d$-, and $f$-wave pairing symmetries depending on the doping level and value of the nearest-neighbor Hubbard repulsion $V$~\cite{Wolf2022}, with chiral $d$-wave pairing becoming dominant when $V$ is small. 
We also find that the strength of the chiral $d$-wave pairing correlations increases as the doping level decreases, consistent with observed gap magnitudes shown in Fig~\ref{fig:1}{\bf e}. 

Our experimental and theoretical results are so far consistent with chiral $d$-wave pairing. Such a pairing state should be characterized by a topological invariant~\cite{VolovikJETP1997}, given by the Chern number 2~\cite{LaughlinPRL1998}. Given the nontrivial topology of the bulk pairing state, we expect the presence of chiral edge modes on sample boundaries~\cite{SenthilPRB1999}, but do not expect unpaired zero energy Majorana bound states at the center of a vortex core, which arise in the case of effectively  spinless $p+\mathrm{i}p$ pairing~\cite{ReadPRB2000}. Indeed, we find no evidence for a Majorana zero mode at the center of a vortex core (see Extended Data Fig.~\ref{fig:vortex}). We did, however, find evidence consistent with the presence of chiral edge modes, as summarized in Fig.\ref{fig:4}\textbf{b}-{\bf i}. 

Figure ~\ref{fig:4}{\bf b} shows a topographic map near a domain wall (vertical dashed line) between the superconducting $(\sqrt{3}\times\sqrt{3})$ domain on the right and semiconducting $(2\sqrt{3}\times2\sqrt{3})$ domain (Ref.~\citenum{MingNC2017}) on the left. Figs.~\ref{fig:4}{\bf c}-{\bf d} show registry-aligned zero-bias conductance maps over the same region of the sample at a function of temperature. The data reveal a region of increased conductance that penetrates the $(\sqrt{3}\times\sqrt{3})$ domain, consistent with the presence of an edge state, which grows as the temperature increases. To estimate the penetration length, we averaged the zero-bias conductance along the vertical direction and plotted it as a function of distance towards the interior, as shown in Fig.~\ref{fig:4}{\bf f}. The data in the superconducting state are reasonably well described by an exponential decay (solid black lines), and we estimate that the decay length grows from $\approx 3.8$ nm at $T = 0.5$~K to $\approx 12.8$ nm at $6$~K. 

The increased zero-bias conductance can also be seen in the individual $dI/dV$ spectra [Fig.~\ref{fig:4}{\bf h}] taken along the line approaching the domain boundary. To highlight the DOS contribution emanating from the edge, we subtracted a $dI/dV$ spectrum taken from deep in bulk from the spectra in {\bf h}, as shown in Fig.~\ref{fig:4}{\bf i}. The data show that the superconducting coherence peaks shift to lower energies when the edge is approached, as indicated by the formation of the peak-dip structures at about $\pm 2.2$~mV in the subtracted spectra. At the same time, the spectral weight in the superconducting gap increases almost uniformly as a function of energy. This behavior is consistent with an overlapping DOS from a linearly dispersing edge mode. To confirm this, we computed the chiral edge modes from a simple chiral $d$-wave mean-field model defined on a cylinder with open boundary conditions (see methods and Extended Data Fig.~\ref{fig:edgemode}). The simulated subtracted data, analogous to panel {\bf i} are shown in Fig.~\ref{fig:4}{\bf j} where they qualitatively reproduce the experimental data. We note, however, that our non-interacting model has some additional substructure in the DOS for the edge-state, which arise from the spatial structure of the edge-state wave function. While we also observe substructure in the experimental spectra, the details differ somewhat from the theory. We expect that a detailed analysis of these features will require the inclusion of interactions in our edge state model, which we leave for future work. While the enhanced zero-bias edge conductance is consistent with the presence of an edge mode, it is difficult to disentangle possible contributions from traditional coherence length effects (or inverse proximity effects~\cite{DeGennesRMP1964}). Nor does the evidence imply that the edge state conductance is topological in nature, which hopefully will be elucidated in future studies. 

 To summarize, our combined results suggest the possibility that the hole-doped Sn/Si(111) Mott insulator system is an unconventional chiral $d$-wave superconductor. This system exhibits a remarkable structural simplicity, with nearly perfect and strictly two-dimensional $(\sqrt{3}\times\sqrt{3})$-Sn layers. It offers the ability to create, access, and characterize well-defined defect states. These aspects allowed us to identify the presence and origin of time-reversal symmetry breaking in the superconducting state. From the applied perspective, this exotic physics is accessible on a simple Si template, which can easily be altered or engineered using standard semiconductor processing or surface science approaches. Modulation-doped semiconductor surfaces and interfaces thus appear to be a promising test bed for studying and exploiting strongly correlated topological states of matter.

\section*{Methods}
\noindent\textbf{Sample preparation}. The hole-doped ($\sqrt{3}\times \sqrt{3}$)-Sn structure was grown on three heavily boron-doped $p$-type silicon substrates with nominal room-temperature resistivities of 0.002, 0.005, and 0.008 $\Omega\cdot\mathrm{cm}$. They correspond to surface hole-doping concentrations of 10\%, 8\% and 6\%, respectively, due to differences in the amount of charge transfer from the bulk to the surface~\cite{MingPRL2017}.  These substrates were annealed to 1200$^\circ$C in ultrahigh vacuum so as to prepare atomically clean Si surfaces.  Sn atoms were deposited onto the clean surface from a thermal effusion cell while keeping the substrate temperature at around 600$^\circ$C. This procedure resulted in the formation of coexisting ($\sqrt{3}\times \sqrt{3}$)-Sn and (2$\sqrt{3}\times 2\sqrt{3}$)-Sn domains. The maximum size (without internal domain boundary) of the ($\sqrt{3}\times \sqrt{3}$)-Sn superconducting domains on each substrate exceeds $200\times 200$~nm$^2$. Small amount additional Sn atoms are deposited onto the surface when the sample is at 120 K, followed by a fast transfer to the STM measurement stage at lower temperature. Additional details can be found in Ref.~\citenum{MingPRL2017}.\\

\noindent\textbf{Scanning tunneling microscopy/spectroscopy measurements}. STM data were acquired using a cryogenic STM (Unisoku) that can cool the sample and tip to 400 mK in the presence of a perpendicular magnetic field of up to 15 T. Differential conductance spectra $dI/dV$ or their spatial maps $g({\bf r},V)$ were acquired using lock-in detection with a typical modulation voltage of $V_\mathrm{rms} = 0.14$~mV and a typical modulation frequency of 673 Hz. A typical $g({\bf r},V)$ map consists of 272$\times$272 pixels measured over a 56$\times$56 nm$^2$ surface area. QPI images are then produced by calculating the power spectral density of the Fourier transforms of the real space conductance map $|g({\bf r},V)|$. \\

\noindent\textbf{STS Fits}. We fit the normalized $dI/dV$ spectra in Fig.~\ref{fig:2}{\bf a} with density of states (DOS)  $N_s(\omega)/N_n(\omega)$, where 
$N_{s (n)}(\omega) = -\frac{2}{\pi N}\sum_{\bf k} \mathrm{Im}G_{11}({\bf k},\omega)$ denotes the DOS in the superconducting (normal) state. Here, 
\begin{equation}\label{eq:Gkw}
    \hat{G}({\bf k},\omega) = 
    \frac{\left(\omega + \mathrm{i}\Gamma\right)\hat{\tau}_0 +  \epsilon({\bf k})\hat{\tau}_3 + \hat{\Delta}(\bf k)}
    {\left(\omega + \mathrm{i}\Gamma\right)^2 - \epsilon^2({\bf k}) - |\Delta({\bf k})|^2},
\end{equation}
is the non-interacting Green's function in Nambu space, $\hat{\tau}_\alpha$ are Pauli matrices, $\epsilon({\bf k})$ is the bare band dispersion, and $\hat{\Delta}({\bf k})$ parameterizes the superconducting gap function in Nambu space $\psi_{\bf k} = [c^{\phantom\dagger}_{{\bf k},\uparrow}, c^{\dagger}_{-{\bf k},\downarrow}]^T$. Note, for the chiral $p$-wave case, the space corresponds to the vector $\vec{d} = \hat{z}$.

To model the Sn surface state, we adopted the tight-binding model derived in Ref.~\citenum{LiNatComm2013}, which is derived from $\it{ab~initio}$ electronic structure calculations. 
The band dispersion is 
\begin{widetext}
\begin{align}\nonumber
    \epsilon({\bf k})=&-2t_1\left[\cos\left(k_xa\right)+2\cos\left(\tfrac{\sqrt{3}}{2}k_ya\right)\cos\left(\tfrac{1}{2}k_xa\right)\right] -2t_2\left[\cos\left(\sqrt{3}k_ya\right)+2\cos\left(\tfrac{3}{2}k_xa\right)
     \cos\left(\tfrac{\sqrt{3}}{2}k_ya\right)\right] \\\nonumber
    &-2t_3\left[\cos\left(2k_xa\right)+2\cos\left(k_xa\right)\cos\left(\sqrt{3}k_ya\right)\right]\\\nonumber
    &-4t_4\left[\cos\left(\tfrac{5}{2}k_xa\right)\cos\left(\tfrac{\sqrt{3}}{2}k_ya\right) +\cos\left(2k_xa\right)\cos\left(\sqrt{3}k_ya\right)+\cos\left(\tfrac{1}{2}k_xa\right)\cos\left(3\tfrac{\sqrt{3}}{2}k_ya\right)\right]  \\
    & -2t_5\left[\cos\left(2\sqrt{3}k_ya\right) + 2\cos\left(3k_xa\right)\cos\left(\sqrt{3}k_ya\right)\right]-\mu,
\label{bareband}\end{align}
with $t_1=52.7$ meV, $t_2 = 0.3881t_1$, $t_3 = 0.1444t_1$, $t_4 = -0.0228t_1$, $t_5 = -0.0318t_1$, and $\mu = -0.017$. Note that we adjusted the chemical potential to put the van Hove singularity $\approx 7.1$ meV below $E_\mathrm{F}$ ~\cite{MingPRL2017}.

To model the normal state, we set $\hat{\Delta}({\bf k}) = 0$. To model an  $s$-wave superconductor, we set $\hat{\Delta}({\bf k}) = \Delta_0\tau_1$. To model the chiral $p$- and $d$-wave cases, we set 
$\hat{\Delta}({\bf k})=\frac{\Delta_l({\bf k})}{2}\left(\hat{\tau}_1+\mathrm{i}\hat{\tau}_2\right) + \frac{\Delta_l^*({\bf k})}{2}\left(\hat{\tau}_1 - \mathrm{i}\hat{\tau}_2\right)$, where $\Delta_{l}({\bf k}) =
2\Delta_0\left[\beta^\prime_l({\bf k}) + \mathrm{i}\beta^{\prime\prime}_l({\bf k})\right]$ and~\cite{ZhouPRL2008} 
\begin{equation*}
    \beta^\prime_{l=1}= \sqrt{3}\sin\left(\tfrac{\sqrt{3}k_ya}{2}\right)\cos\left(\tfrac{k_xa}{2}\right)\quad\mathrm{and}\quad
    \beta^{\prime\prime}_{l=1}=\sin\left(k_x\right)+\cos\left(\tfrac{\sqrt{3}k_ya}{2}\right)\sin\left(\tfrac{k_xa}{2}\right)
\end{equation*}
for chiral $p$-wave pairing and 
\begin{equation*}
    \beta^\prime_{l=2}=\cos\left(k_xa\right)-\cos\left(\tfrac{\sqrt{3}k_ya}{2}\right)\cos\left(\tfrac{k_xa}{2}\right)\quad\mathrm{and}\quad 
    \beta_{l=2}^{\prime\prime}= \sqrt{3}\sin\left(\tfrac{\sqrt{3}k_ya}{2}\right)\sin\left(\tfrac{k_xa}{2}\right)
\end{equation*}
for chiral $d$-wave pairing. We have also fit the spectra with nematic $d$-wave order parameters (see Supplementary Note 2) and found that they are unable to reproduce the experimental spectra.
\end{widetext}

We then fit theoretical DOS to the STS data treating $\Delta_0$ and $\Gamma$ as fitting parameters. The best fits shown in Fig.~\ref{fig:2}{\bf a} are obtained with $(\Delta_0,\Gamma) = (1.88,0.29)$, $(0.76,0.081)$, and $(0.54,0.17)$ for the $s$-, chiral $p$-, and chiral $d$-wave cases, respectively, in units of meV. \\

\noindent\textbf{Quasiparticle interference calculations}. The QPI spectra are calculated using the Born approximation to the $T$-matrix formalism, with a single point-like potential located at the origin. The Fourier transform of the modulation in the electron density is given by 
\begin{align}
    \delta N({\bf q},\omega) =\frac{1}{N}\sum_{{\bf k}}\mathrm{Tr}\left[\mathrm{Im}\left\{\frac{\hat{\tau}_0+\hat{\tau}_3}{2\pi}\hat{G}({\bf k},\omega)\hat{V}\hat{G}({\bf p},\omega)\right\}\right], 
\end{align}
where ${\bf p} = {\bf k}+{\bf q}$, $\hat{G}({\bf k},\omega)$ is the Green's function given by Eq.~\eqref{eq:Gkw}, $\hat{V}$ is the impurity potential, and $\epsilon({\bf k})$ and $\Delta({\bf k})$ are the bare band dispersion and superconducting order parameter, respectively. We considered both magnetic ($\hat{V} = V_0\tau_0$) and nonmagnetic ($\hat{V} = V_0\tau_3$) scatterers, whose strengths are parameterized by $V_0$. For the superconducting gap, we considered many different pairing symmetries, as parameterized above (see Extended Data Figure \ref{fig:extended_QPI}).\\

\noindent\textbf{DFT calculations}.
Plane-wave DFT calculations were implemented using the Quantum Espresso open source computer code~\cite{GiannozziJPCM2009}. We have used the PBE exchange-correlation functional~\cite{PerdewPRL1996} and the ultrasoft pseudopotentials provided by the code~\cite{GoedeckerPRB1996, HartwigsenPRB1998}. The energy cutoff for the plane waves is 40 Ry. While DFT cannot capture the Mott state of the ($\sqrt{3}\times \sqrt{3}$)-Sn system, it accurately captures the ground state structure~\cite{PerezPRL2001}. We employed a ($9 \times 9$) supercell with 6 Si layers and 27 Sn adatoms in $T_4$ positions. In total, there are 594 atoms in the unit cell. To simulate the substitutional Si defect, we replaced one of the 27 Sn adatoms with a Si atom. Based on the experimental results, the interstitial Sn adatom is placed at the center of an equilateral triangle formed by three adjacent Sn atoms of the original ($\sqrt{3}\times \sqrt{3}$)-Sn surface. Hence the total number of Sn atoms is 28; see Extended Data Fig.~\ref{fig:Defect_and_Simulation}. The bottom two Si layers and the H layer are fixed in these simulations. The ($9 \times 9$) first Brillouin zone was sampled with a $2\times 2$ Monkhorst-Pack (MP) grid~\cite{MonkhorstPRB1976} and the geometry is relaxed until the forces are lower than 0.001 Ry/Bohr. In the total-energy minimized geometry, the Sn atoms forming the triangle are located $\sim 0.2$ \AA{} above the atoms of the ($\sqrt{3}\times \sqrt{3}$)-Sn layer. The additional Sn adatom at the center of the triangle is located $\sim 0.11$ \AA{} above the ($\sqrt{3}\times \sqrt{3}$)-Sn layer, i.e., $\sim 0.09$ \AA{} below its nearest neighbors.\\

\noindent\textbf{STM image simulations}.
The simulated STM images for both defect structures were calculated using the Keldysh-Green Function formalism~\cite{BlancoPRB2004} together with the Fireball local-orbital DFT Hamiltonian ~\cite{LewisPSSB2011}. This procedure has been successfully used in many works before (e.g. see Ref.~\citenum{GonzalezPRL2004}). In our simulations, a standard W tip is placed at a distance of 5 \AA{} above the surface and images were generated for tunneling parameters close to the experimental conditions. In Extended Data Fig.~\ref{fig:Defect_and_Simulation} we show the experimental and simulated images side by side, showing excellent agreement for the chosen tunneling parameters. The STM images can be correlated to the different atomic heights and  Projected Density of States (PDOS) on the different atoms. The PDOS on the interstitial Sn adatom, calculated with the local-orbital DFT code, is shown in Extended Data Fig.~\ref{fig:Defect_and_Simulation}. A similar procedure was implemented for the substitutional Si defect. The corresponding experimental and theoretical images are shown in Extended Data Fig.~\ref{fig:Defect_and_Simulation} and are in good agreement. \\ 

\noindent{\bf DCA calculations}. 
DCA calculations for the triangular lattice single-band Hubbard model were performed with the DCA++ package~\cite{UrsCompPhysComm2020} using a continuous-time quantum Monte Carlo impurity solver. Here, we consider a $N = 3\times 3$ cluster of Sn atoms with the bare band structure as given in Ref.~\citenum{LiNatComm2013} but including only up to third nearest neighbor hopping (i.e. $t_4 = t_5= 0$), since the cluster cannot support longer range hopping. We further set $U = 0.66$ eV~\cite{LiNatComm2013}. The DCA calculations measured the single-particle Green's function and two-particle Green’s function in the particle-particle channel with zero momentum and frequency transfer. From that, we extracted the irreducible particle-particle vertex and then solved the Bethe-Salpeter equation (BSE) to obtain the leading eigenvalues and eigenvectors in the particle-particle channel, as described in Ref.~\citenum{MaierPRL2006}. The pairing symmetry of the dominant superconducting instability can be extracted from the leading eigenvector. \\

\noindent{\bf Edge state calculations.} 
We compute the topological edge states of the chiral $d+\mathrm{i}d$ superconductor by solving the same triangular lattice model defined in Eq.~\eqref{bareband} but defined on a cluster with open boundary conditions in the $y$ direction and periodic boundary condition in the $x$ direction, as shown in Extended Data Fig.~\ref{fig:edgemode}. Introducing the labels $m,n$ for the chains stacked in the $y$ direction, and $k$ for the momentum along the edge, the Hamiltonian can be written as
\begin{equation}
H = \frac12 \sum_{mn}\sum_{ k} \Phi^\dagger_{m}({ k}) \mathcal H^{\phantom\dagger}_{mn}(k)\Phi^{\phantom\dagger}_{n}({ k}), 
\end{equation}
where $\Phi^\dagger_{n}(k)= [c^\dagger_{n\uparrow}({k}), c^{\phantom\dagger}_{n\downarrow}(-{k})]$ is a Nambu spinor collecting electron and hole degree of freedom. The Hamiltonian matrix $\mathcal{H}_{mn}$ describes how chains $m$ and $n$ are coupled. It is convenient to decompose $\mathcal H_{mn}$ as (momentum dependence is suppressed)
\begin{equation}
\mathcal H_{mn} = \sum_{p=-l}^l \mathcal H_p \delta_{p,n-m},
\end{equation}
where $\mathcal H_p$ captures the coupling of $|p|$-th nearest neighbor chains and satisfies $\mathcal H_{-p}=\mathcal H^\dagger_p$. Here $l$ denotes the extent of the coupling; if only first nearest neighbor chains are coupled $l=1$. In the present problem $l=4$. The matrices $\mathcal H_p$ are given by
\begin{equation}
\mathcal H_p = \begin{pmatrix} \epsilon_p &  \Delta_p \\  \Delta^*_{-p} & -\epsilon_p \end{pmatrix}.
\end{equation}

The Hamiltonian defines a Schr\"odinger equation given by
\begin{equation}
 \sum_{p=-4}^4 \mathcal H_p({ k}) \Psi_{n+p}({ k}) = E  \Psi_{n}({ k}), 
\end{equation}
where $\Psi_{n}({ k})=[u_{n }({k}), v_{n }({ k})]^T$ is the wave function with electron and hole components $u_{n }({ k}) $ and $v_{n }({ k}) $. We solve this system of equations numerically on a finite slab to obtain the quasiparticle spectrum $E^\xi_{k}$ in the presence of an edge. Here $\xi$ is an index to label eigenvalues at given ${ k}$. For a system with $N$ chains one has $\xi=1,\ldots,2N$.

Finally, we compute the chain-resolved local density of states $N(n,\omega)$ as
\begin{equation}
N(n,\omega) =  \sum_{{ k},\xi} |u^\xi_{n }({ k}) |^2 \delta( \omega - E^\xi_{{}})+ |v^\xi_{n }({ k}) |^2\delta( \omega + E^\xi_{k}).
\end{equation}

\noindent{\bf Acknowledgments}: We thank C.~D. Batista, P.~J. Hirschfeld, P. Kent, A. Tennant, and R. Zhang for fruitful discussions. The experimental work and QPI calculations were supported by the Guangdong Basic and Applied Basic Research Foundation, Ref No. 2021A1515012034 and by the Office of Naval Research under Grant No. N00014-18-1-2675. F.~M. acknowledges the support by NSFC (No. 12174456) and the Guangdong Basic and Applied Basic Research Foundation (grants no. 2020B1515020009) C.~G. acknowledges financial support from the Community of Madrid through the project NANOMAGCOST CM-PS2018/NMT-4321 and the computer resources at Altamira, and the technical support provided by the Instituto de Física de Cantabria (IFCA), Project: QHS-2021-3-0005. J.~O. acknowledges financial support by the Spanish Ministry of Science and Innovation through grants MAT2017-88258-R and CEX2018-000805-M (Mar{\'i}a de Maeztu Programme for Units of Excellence in R\&D).  The DCA calculations were supported by the Scientific Discovery through Advanced Computing (SciDAC) program funded by the U.S. Department of Energy, Office of Science, Advanced Scientific Computing Research and Basic Energy Sciences, Division of Materials Sciences and Engineering. This research also used resources of the Oak Ridge Leadership Computing Facility, which is a DOE Office of Science User Facility supported under Contract DE-AC05-00OR22725. \\

\noindent {\bf Data availability}: The data supporting this study will be made available upon request. \\

\noindent{\bf Code availability}: The DCA++ code used for this project can be obtained at \url{https://github.com/ CompFUSE/DCA}. The Quantum Espresso code can be obtained from \url{https://www.quantum-espresso.org/}. The Fireball code can be obtained from \url{https://github.com/fireball-QMD}. The other codes supporting this study will be made available upon request. \\

\noindent{\bf Author contributions}: F.~M., X.~W. contributed equally to this work. F.~M., X.~W., C.~C and K.~D.~W. prepared the samples and performed the STM experiments. P.~M. and T.~A.~M. performed the DCA calculations. J.~S. and J.~W.~F.~V.  performed edge state calculations. C.~G. and J.~O. performed the DFT calculations and STM image simulations. S.~J. performed the QPI calculations. S.~J. and H.~H.~W. conceived and supervised the project, and wrote the manuscript with input from all authors. \\

\noindent{\bf Competing interests}: The authors declare that they have no competing financial interests. \\

\noindent \textbf{Correspondence} and requests for materials should be addressed to K.D.~Wang (email: \href{wangkd@sustech.edu.cn}{wangkd@sustech.edu.cn}), S.J. (email: \href{sjohn145@utk.edu}{sjohn145@utk.edu}), or H.H.W. (email: \href{hanno@utk.edu}{hanno@utk.edu}).

\bibliography{arXiv_references}

\begin{thebibliography}{58}%
\makeatletter
\providecommand \@ifxundefined [1]{%
 \@ifx{#1\undefined}
}%
\providecommand \@ifnum [1]{%
 \ifnum #1\expandafter \@firstoftwo
 \else \expandafter \@secondoftwo
 \fi
}%
\providecommand \@ifx [1]{%
 \ifx #1\expandafter \@firstoftwo
 \else \expandafter \@secondoftwo
 \fi
}%
\providecommand \natexlab [1]{#1}%
\providecommand \enquote  [1]{``#1''}%
\providecommand \bibnamefont  [1]{#1}%
\providecommand \bibfnamefont [1]{#1}%
\providecommand \citenamefont [1]{#1}%
\providecommand \href@noop [0]{\@secondoftwo}%
\providecommand \href [0]{\begingroup \@sanitize@url \@href}%
\providecommand \@href[1]{\@@startlink{#1}\@@href}%
\providecommand \@@href[1]{\endgroup#1\@@endlink}%
\providecommand \@sanitize@url [0]{\catcode `\\12\catcode `\$12\catcode
  `\&12\catcode `\#12\catcode `\^12\catcode `\_12\catcode `\%12\relax}%
\providecommand \@@startlink[1]{}%
\providecommand \@@endlink[0]{}%
\providecommand \url  [0]{\begingroup\@sanitize@url \@url }%
\providecommand \@url [1]{\endgroup\@href {#1}{\urlprefix }}%
\providecommand \urlprefix  [0]{URL }%
\providecommand \Eprint [0]{\href }%
\providecommand \doibase [0]{https://doi.org/}%
\providecommand \selectlanguage [0]{\@gobble}%
\providecommand \bibinfo  [0]{\@secondoftwo}%
\providecommand \bibfield  [0]{\@secondoftwo}%
\providecommand \translation [1]{[#1]}%
\providecommand \BibitemOpen [0]{}%
\providecommand \bibitemStop [0]{}%
\providecommand \bibitemNoStop [0]{.\EOS\space}%
\providecommand \EOS [0]{\spacefactor3000\relax}%
\providecommand \BibitemShut  [1]{\csname bibitem#1\endcsname}%
\let\auto@bib@innerbib\@empty
\bibitem [{\citenamefont {Bardeen}\ \emph {et~al.}(1957)\citenamefont
  {Bardeen}, \citenamefont {Cooper},\ and\ \citenamefont {Schrieffer}}]{BCS}%
  \BibitemOpen
  \bibfield  {author} {\bibinfo {author} {\bibfnamefont {J.}~\bibnamefont
  {Bardeen}}, \bibinfo {author} {\bibfnamefont {L.~N.}\ \bibnamefont
  {Cooper}},\ and\ \bibinfo {author} {\bibfnamefont {J.~R.}\ \bibnamefont
  {Schrieffer}},\ }\bibfield  {title} {\bibinfo {title} {Theory of
  superconductivity},\ }\href {https://doi.org/10.1103/PhysRev.108.1175}
  {\bibfield  {journal} {\bibinfo  {journal} {Phys. Rev.}\ }\textbf {\bibinfo
  {volume} {108}},\ \bibinfo {pages} {1175} (\bibinfo {year}
  {1957})}\BibitemShut {NoStop}%
\bibitem [{\citenamefont {Kohn}\ and\ \citenamefont
  {Luttinger}(1965)}]{KohnPRL1965}%
  \BibitemOpen
  \bibfield  {author} {\bibinfo {author} {\bibfnamefont {W.}~\bibnamefont
  {Kohn}}\ and\ \bibinfo {author} {\bibfnamefont {J.~M.}\ \bibnamefont
  {Luttinger}},\ }\bibfield  {title} {\bibinfo {title} {New mechanism for
  superconductivity},\ }\href {https://doi.org/10.1103/PhysRevLett.15.524}
  {\bibfield  {journal} {\bibinfo  {journal} {Phys. Rev. Lett.}\ }\textbf
  {\bibinfo {volume} {15}},\ \bibinfo {pages} {524} (\bibinfo {year}
  {1965})}\BibitemShut {NoStop}%
\bibitem [{\citenamefont {Scalapino}(2012)}]{ScalapinoRMP2012}%
  \BibitemOpen
  \bibfield  {author} {\bibinfo {author} {\bibfnamefont {D.~J.}\ \bibnamefont
  {Scalapino}},\ }\bibfield  {title} {\bibinfo {title} {A common thread: The
  pairing interaction for unconventional superconductors},\ }\href
  {https://doi.org/10.1103/RevModPhys.84.1383} {\bibfield  {journal} {\bibinfo
  {journal} {Rev. Mod. Phys.}\ }\textbf {\bibinfo {volume} {84}},\ \bibinfo
  {pages} {1383} (\bibinfo {year} {2012})}\BibitemShut {NoStop}%
\bibitem [{\citenamefont {Tsuei}\ and\ \citenamefont
  {Kirtley}(2000)}]{TsueiRMP2000}%
  \BibitemOpen
  \bibfield  {author} {\bibinfo {author} {\bibfnamefont {C.~C.}\ \bibnamefont
  {Tsuei}}\ and\ \bibinfo {author} {\bibfnamefont {J.~R.}\ \bibnamefont
  {Kirtley}},\ }\bibfield  {title} {\bibinfo {title} {Pairing symmetry in
  cuprate superconductors},\ }\href {https://doi.org/10.1103/RevModPhys.72.969}
  {\bibfield  {journal} {\bibinfo  {journal} {Rev. Mod. Phys.}\ }\textbf
  {\bibinfo {volume} {72}},\ \bibinfo {pages} {969} (\bibinfo {year}
  {2000})}\BibitemShut {NoStop}%
\bibitem [{\citenamefont {Read}\ and\ \citenamefont
  {Green}(2000)}]{ReadPRB2000}%
  \BibitemOpen
  \bibfield  {author} {\bibinfo {author} {\bibfnamefont {N.}~\bibnamefont
  {Read}}\ and\ \bibinfo {author} {\bibfnamefont {D.}~\bibnamefont {Green}},\
  }\bibfield  {title} {\bibinfo {title} {Paired states of fermions in two
  dimensions with breaking of parity and time-reversal symmetries and the
  fractional quantum {H}all effect},\ }\href
  {https://doi.org/10.1103/PhysRevB.61.10267} {\bibfield  {journal} {\bibinfo
  {journal} {Phys. Rev. B}\ }\textbf {\bibinfo {volume} {61}},\ \bibinfo
  {pages} {10267} (\bibinfo {year} {2000})}\BibitemShut {NoStop}%
\bibitem [{\citenamefont {Joynt}\ and\ \citenamefont
  {Taillefer}(2002)}]{JoyntRMP2002}%
  \BibitemOpen
  \bibfield  {author} {\bibinfo {author} {\bibfnamefont {R.}~\bibnamefont
  {Joynt}}\ and\ \bibinfo {author} {\bibfnamefont {L.}~\bibnamefont
  {Taillefer}},\ }\bibfield  {title} {\bibinfo {title} {The superconducting
  phases of {${\mathrm{UPt}}_{3}$}},\ }\href
  {https://doi.org/10.1103/RevModPhys.74.235} {\bibfield  {journal} {\bibinfo
  {journal} {Rev. Mod. Phys.}\ }\textbf {\bibinfo {volume} {74}},\ \bibinfo
  {pages} {235} (\bibinfo {year} {2002})}\BibitemShut {NoStop}%
\bibitem [{\citenamefont {Nandkishore}\ \emph {et~al.}(2012)\citenamefont
  {Nandkishore}, \citenamefont {Levitov},\ and\ \citenamefont
  {Chubukov}}]{NandkishoreNatPhys2012}%
  \BibitemOpen
  \bibfield  {author} {\bibinfo {author} {\bibfnamefont {R.}~\bibnamefont
  {Nandkishore}}, \bibinfo {author} {\bibfnamefont {L.~S.}\ \bibnamefont
  {Levitov}},\ and\ \bibinfo {author} {\bibfnamefont {A.~V.}\ \bibnamefont
  {Chubukov}},\ }\bibfield  {title} {\bibinfo {title} {Chiral superconductivity
  from repulsive interactions in doped graphene},\ }\href
  {https://doi.org/10.1038/nphys2208} {\bibfield  {journal} {\bibinfo
  {journal} {Nature Physics}\ }\textbf {\bibinfo {volume} {8}},\ \bibinfo
  {pages} {158} (\bibinfo {year} {2012})}\BibitemShut {NoStop}%
\bibitem [{\citenamefont {Kallin}(2012)}]{Kallin2012}%
  \BibitemOpen
  \bibfield  {author} {\bibinfo {author} {\bibfnamefont {C.}~\bibnamefont
  {Kallin}},\ }\bibfield  {title} {\bibinfo {title} {Chiral $p$-wave order in
  {Sr$_2${RuO}$_4$}},\ }\href {https://doi.org/10.1088/0034-4885/75/4/042501}
  {\bibfield  {journal} {\bibinfo  {journal} {Reports on Progress in Physics}\
  }\textbf {\bibinfo {volume} {75}},\ \bibinfo {pages} {042501} (\bibinfo
  {year} {2012})}\BibitemShut {NoStop}%
\bibitem [{\citenamefont {Black-Schaffer}(2012)}]{Black-SchafferPRL2012}%
  \BibitemOpen
  \bibfield  {author} {\bibinfo {author} {\bibfnamefont {A.~M.}\ \bibnamefont
  {Black-Schaffer}},\ }\bibfield  {title} {\bibinfo {title} {Edge properties
  and {M}ajorana fermions in the proposed chiral $d$-wave superconducting state
  of doped graphene},\ }\href {https://doi.org/10.1103/PhysRevLett.109.197001}
  {\bibfield  {journal} {\bibinfo  {journal} {Phys. Rev. Lett.}\ }\textbf
  {\bibinfo {volume} {109}},\ \bibinfo {pages} {197001} (\bibinfo {year}
  {2012})}\BibitemShut {NoStop}%
\bibitem [{\citenamefont {Kiesel}\ \emph {et~al.}(2013)\citenamefont {Kiesel},
  \citenamefont {Platt}, \citenamefont {Hanke},\ and\ \citenamefont
  {Thomale}}]{KieselPRL2013}%
  \BibitemOpen
  \bibfield  {author} {\bibinfo {author} {\bibfnamefont {M.~L.}\ \bibnamefont
  {Kiesel}}, \bibinfo {author} {\bibfnamefont {C.}~\bibnamefont {Platt}},
  \bibinfo {author} {\bibfnamefont {W.}~\bibnamefont {Hanke}},\ and\ \bibinfo
  {author} {\bibfnamefont {R.}~\bibnamefont {Thomale}},\ }\bibfield  {title}
  {\bibinfo {title} {Model evidence of an anisotropic chiral
  $d\mathbf{+}id$-wave pairing state for the water-intercalated
  {${\mathrm{Na}}_{x}{\mathrm{CoO}}_{2}\ifmmode\cdot\else\textperiodcentered\fi{}y{\mathrm{H}}_{2}\mathrm{O}$}
  superconductor},\ }\href {https://doi.org/10.1103/PhysRevLett.111.097001}
  {\bibfield  {journal} {\bibinfo  {journal} {Phys. Rev. Lett.}\ }\textbf
  {\bibinfo {volume} {111}},\ \bibinfo {pages} {097001} (\bibinfo {year}
  {2013})}\BibitemShut {NoStop}%
\bibitem [{\citenamefont {Black-Schaffer}\ and\ \citenamefont
  {Honerkamp}(2014)}]{Black-SchafferJPCM2014}%
  \BibitemOpen
  \bibfield  {author} {\bibinfo {author} {\bibfnamefont {A.~M.}\ \bibnamefont
  {Black-Schaffer}}\ and\ \bibinfo {author} {\bibfnamefont {C.}~\bibnamefont
  {Honerkamp}},\ }\bibfield  {title} {\bibinfo {title} {Chiral $d$-wave
  superconductivity in doped graphene},\ }\href
  {https://doi.org/10.1088/0953-8984/26/42/423201} {\bibfield  {journal}
  {\bibinfo  {journal} {Journal of Physics: Condensed Matter}\ }\textbf
  {\bibinfo {volume} {26}},\ \bibinfo {pages} {423201} (\bibinfo {year}
  {2014})}\BibitemShut {NoStop}%
\bibitem [{\citenamefont {Kallin}\ and\ \citenamefont
  {Berlinsky}(2016)}]{KallinRPP2016}%
  \BibitemOpen
  \bibfield  {author} {\bibinfo {author} {\bibfnamefont {C.}~\bibnamefont
  {Kallin}}\ and\ \bibinfo {author} {\bibfnamefont {J.}~\bibnamefont
  {Berlinsky}},\ }\bibfield  {title} {\bibinfo {title} {Chiral
  superconductors},\ }\href {https://doi.org/10.1088/0034-4885/79/5/054502}
  {\bibfield  {journal} {\bibinfo  {journal} {Reports on Progress in Physics}\
  }\textbf {\bibinfo {volume} {79}},\ \bibinfo {pages} {054502} (\bibinfo
  {year} {2016})}\BibitemShut {NoStop}%
\bibitem [{\citenamefont {Mackenzie}\ \emph {et~al.}(2017)\citenamefont
  {Mackenzie}, \citenamefont {Scaffidi}, \citenamefont {Hicks},\ and\
  \citenamefont {Maeno}}]{Mackenzie2017}%
  \BibitemOpen
  \bibfield  {author} {\bibinfo {author} {\bibfnamefont {A.~P.}\ \bibnamefont
  {Mackenzie}}, \bibinfo {author} {\bibfnamefont {T.}~\bibnamefont {Scaffidi}},
  \bibinfo {author} {\bibfnamefont {C.~W.}\ \bibnamefont {Hicks}},\ and\
  \bibinfo {author} {\bibfnamefont {Y.}~\bibnamefont {Maeno}},\ }\bibfield
  {title} {\bibinfo {title} {Even odder after twenty-three years: the
  superconducting order parameter puzzle of {Sr$_2$RuO$_4$}},\ }\href
  {https://doi.org/10.1038/s41535-017-0045-4} {\bibfield  {journal} {\bibinfo
  {journal} {npj Quantum Materials}\ }\textbf {\bibinfo {volume} {2}},\
  \bibinfo {pages} {40} (\bibinfo {year} {2017})}\BibitemShut {NoStop}%
\bibitem [{\citenamefont {Pustogow}\ \emph {et~al.}(2019)\citenamefont
  {Pustogow}, \citenamefont {Luo}, \citenamefont {Chronister}, \citenamefont
  {Su}, \citenamefont {Sokolov}, \citenamefont {Jerzembeck}, \citenamefont
  {Mackenzie}, \citenamefont {Hicks}, \citenamefont {Kikugawa}, \citenamefont
  {Raghu}, \citenamefont {Bauer},\ and\ \citenamefont {Brown}}]{Pustogow2019}%
  \BibitemOpen
  \bibfield  {author} {\bibinfo {author} {\bibfnamefont {A.}~\bibnamefont
  {Pustogow}}, \bibinfo {author} {\bibfnamefont {Y.}~\bibnamefont {Luo}},
  \bibinfo {author} {\bibfnamefont {A.}~\bibnamefont {Chronister}}, \bibinfo
  {author} {\bibfnamefont {Y.~S.}\ \bibnamefont {Su}}, \bibinfo {author}
  {\bibfnamefont {D.~A.}\ \bibnamefont {Sokolov}}, \bibinfo {author}
  {\bibfnamefont {F.}~\bibnamefont {Jerzembeck}}, \bibinfo {author}
  {\bibfnamefont {A.~P.}\ \bibnamefont {Mackenzie}}, \bibinfo {author}
  {\bibfnamefont {C.~W.}\ \bibnamefont {Hicks}}, \bibinfo {author}
  {\bibfnamefont {N.}~\bibnamefont {Kikugawa}}, \bibinfo {author}
  {\bibfnamefont {S.}~\bibnamefont {Raghu}}, \bibinfo {author} {\bibfnamefont
  {E.~D.}\ \bibnamefont {Bauer}},\ and\ \bibinfo {author} {\bibfnamefont
  {S.~E.}\ \bibnamefont {Brown}},\ }\bibfield  {title} {\bibinfo {title}
  {Constraints on the superconducting order parameter in {Sr$_2$RuO$_4$} from
  oxygen-17 nuclear magnetic resonance},\ }\href
  {https://doi.org/10.1038/s41586-019-1596-2} {\bibfield  {journal} {\bibinfo
  {journal} {Nature}\ }\textbf {\bibinfo {volume} {574}},\ \bibinfo {pages}
  {72} (\bibinfo {year} {2019})}\BibitemShut {NoStop}%
\bibitem [{\citenamefont {Jiao}\ \emph {et~al.}(2020)\citenamefont {Jiao},
  \citenamefont {Howard}, \citenamefont {Ran}, \citenamefont {Wang},
  \citenamefont {Rodriguez}, \citenamefont {Sigrist}, \citenamefont {Wang},
  \citenamefont {Butch},\ and\ \citenamefont {Madhavan}}]{JiaoNature2020}%
  \BibitemOpen
  \bibfield  {author} {\bibinfo {author} {\bibfnamefont {L.}~\bibnamefont
  {Jiao}}, \bibinfo {author} {\bibfnamefont {S.}~\bibnamefont {Howard}},
  \bibinfo {author} {\bibfnamefont {S.}~\bibnamefont {Ran}}, \bibinfo {author}
  {\bibfnamefont {Z.}~\bibnamefont {Wang}}, \bibinfo {author} {\bibfnamefont
  {J.~O.}\ \bibnamefont {Rodriguez}}, \bibinfo {author} {\bibfnamefont
  {M.}~\bibnamefont {Sigrist}}, \bibinfo {author} {\bibfnamefont
  {Z.}~\bibnamefont {Wang}}, \bibinfo {author} {\bibfnamefont {N.~P.}\
  \bibnamefont {Butch}},\ and\ \bibinfo {author} {\bibfnamefont
  {V.}~\bibnamefont {Madhavan}},\ }\bibfield  {title} {\bibinfo {title} {Chiral
  superconductivity in heavy-fermion metal {UTe$_2$}},\ }\href
  {https://doi.org/10.1038/s41586-020-2122-2} {\bibfield  {journal} {\bibinfo
  {journal} {Nature}\ }\textbf {\bibinfo {volume} {579}},\ \bibinfo {pages}
  {523} (\bibinfo {year} {2020})}\BibitemShut {NoStop}%
\bibitem [{\citenamefont {Li}\ \emph {et~al.}(2013)\citenamefont {Li},
  \citenamefont {H{\"o}pfner}, \citenamefont {Sch{\"a}fer}, \citenamefont
  {Blumenstein}, \citenamefont {Meyer}, \citenamefont {Bostwick}, \citenamefont
  {Rotenberg}, \citenamefont {Claessen},\ and\ \citenamefont
  {Hanke}}]{LiNatComm2013}%
  \BibitemOpen
  \bibfield  {author} {\bibinfo {author} {\bibfnamefont {G.}~\bibnamefont
  {Li}}, \bibinfo {author} {\bibfnamefont {P.}~\bibnamefont {H{\"o}pfner}},
  \bibinfo {author} {\bibfnamefont {J.}~\bibnamefont {Sch{\"a}fer}}, \bibinfo
  {author} {\bibfnamefont {C.}~\bibnamefont {Blumenstein}}, \bibinfo {author}
  {\bibfnamefont {S.}~\bibnamefont {Meyer}}, \bibinfo {author} {\bibfnamefont
  {A.}~\bibnamefont {Bostwick}}, \bibinfo {author} {\bibfnamefont
  {E.}~\bibnamefont {Rotenberg}}, \bibinfo {author} {\bibfnamefont
  {R.}~\bibnamefont {Claessen}},\ and\ \bibinfo {author} {\bibfnamefont
  {W.}~\bibnamefont {Hanke}},\ }\bibfield  {title} {\bibinfo {title} {Magnetic
  order in a frustrated two-dimensional atom lattice at a semiconductor
  surface},\ }\href {https://doi.org/10.1038/ncomms2617} {\bibfield  {journal}
  {\bibinfo  {journal} {Nature Communications}\ }\textbf {\bibinfo {volume}
  {4}},\ \bibinfo {pages} {1620} (\bibinfo {year} {2013})}\BibitemShut
  {NoStop}%
\bibitem [{\citenamefont {Wu}\ \emph {et~al.}(2020)\citenamefont {Wu},
  \citenamefont {Ming}, \citenamefont {Smith}, \citenamefont {Liu},
  \citenamefont {Ye}, \citenamefont {Wang}, \citenamefont {Johnston},\ and\
  \citenamefont {Weitering}}]{WuPRL2020}%
  \BibitemOpen
  \bibfield  {author} {\bibinfo {author} {\bibfnamefont {X.}~\bibnamefont
  {Wu}}, \bibinfo {author} {\bibfnamefont {F.}~\bibnamefont {Ming}}, \bibinfo
  {author} {\bibfnamefont {T.~S.}\ \bibnamefont {Smith}}, \bibinfo {author}
  {\bibfnamefont {G.}~\bibnamefont {Liu}}, \bibinfo {author} {\bibfnamefont
  {F.}~\bibnamefont {Ye}}, \bibinfo {author} {\bibfnamefont {K.}~\bibnamefont
  {Wang}}, \bibinfo {author} {\bibfnamefont {S.}~\bibnamefont {Johnston}},\
  and\ \bibinfo {author} {\bibfnamefont {H.~H.}\ \bibnamefont {Weitering}},\
  }\bibfield  {title} {\bibinfo {title} {Superconductivity in a hole-doped
  {M}ott-insulating triangular adatom layer on a silicon surface},\ }\href
  {https://doi.org/10.1103/PhysRevLett.125.117001} {\bibfield  {journal}
  {\bibinfo  {journal} {Phys. Rev. Lett.}\ }\textbf {\bibinfo {volume} {125}},\
  \bibinfo {pages} {117001} (\bibinfo {year} {2020})}\BibitemShut {NoStop}%
\bibitem [{\citenamefont {Ming}\ \emph
  {et~al.}(2017{\natexlab{a}})\citenamefont {Ming}, \citenamefont {Johnston},
  \citenamefont {Mulugeta}, \citenamefont {Smith}, \citenamefont {Vilmercati},
  \citenamefont {Lee}, \citenamefont {Maier}, \citenamefont {Snijders},\ and\
  \citenamefont {Weitering}}]{MingPRL2017}%
  \BibitemOpen
  \bibfield  {author} {\bibinfo {author} {\bibfnamefont {F.}~\bibnamefont
  {Ming}}, \bibinfo {author} {\bibfnamefont {S.}~\bibnamefont {Johnston}},
  \bibinfo {author} {\bibfnamefont {D.}~\bibnamefont {Mulugeta}}, \bibinfo
  {author} {\bibfnamefont {T.~S.}\ \bibnamefont {Smith}}, \bibinfo {author}
  {\bibfnamefont {P.}~\bibnamefont {Vilmercati}}, \bibinfo {author}
  {\bibfnamefont {G.}~\bibnamefont {Lee}}, \bibinfo {author} {\bibfnamefont
  {T.~A.}\ \bibnamefont {Maier}}, \bibinfo {author} {\bibfnamefont {P.~C.}\
  \bibnamefont {Snijders}},\ and\ \bibinfo {author} {\bibfnamefont {H.~H.}\
  \bibnamefont {Weitering}},\ }\bibfield  {title} {\bibinfo {title}
  {Realization of a hole-doped {M}ott insulator on a triangular silicon
  lattice},\ }\href {https://doi.org/10.1103/PhysRevLett.119.266802} {\bibfield
   {journal} {\bibinfo  {journal} {Phys. Rev. Lett.}\ }\textbf {\bibinfo
  {volume} {119}},\ \bibinfo {pages} {266802} (\bibinfo {year}
  {2017}{\natexlab{a}})}\BibitemShut {NoStop}%
\bibitem [{\citenamefont {Lee}\ \emph {et~al.}(2006)\citenamefont {Lee},
  \citenamefont {Nagaosa},\ and\ \citenamefont {Wen}}]{LeeRMP2006}%
  \BibitemOpen
  \bibfield  {author} {\bibinfo {author} {\bibfnamefont {P.~A.}\ \bibnamefont
  {Lee}}, \bibinfo {author} {\bibfnamefont {N.}~\bibnamefont {Nagaosa}},\ and\
  \bibinfo {author} {\bibfnamefont {X.-G.}\ \bibnamefont {Wen}},\ }\bibfield
  {title} {\bibinfo {title} {Doping a {M}ott insulator: Physics of
  high-temperature superconductivity},\ }\href
  {https://doi.org/10.1103/RevModPhys.78.17} {\bibfield  {journal} {\bibinfo
  {journal} {Rev. Mod. Phys.}\ }\textbf {\bibinfo {volume} {78}},\ \bibinfo
  {pages} {17} (\bibinfo {year} {2006})}\BibitemShut {NoStop}%
\bibitem [{\citenamefont {Cao}\ \emph {et~al.}(2018)\citenamefont {Cao},
  \citenamefont {Ayral}, \citenamefont {Zhong}, \citenamefont {Parcollet},
  \citenamefont {Manske},\ and\ \citenamefont {Hansmann}}]{CaoPRB2018}%
  \BibitemOpen
  \bibfield  {author} {\bibinfo {author} {\bibfnamefont {X.}~\bibnamefont
  {Cao}}, \bibinfo {author} {\bibfnamefont {T.}~\bibnamefont {Ayral}}, \bibinfo
  {author} {\bibfnamefont {Z.}~\bibnamefont {Zhong}}, \bibinfo {author}
  {\bibfnamefont {O.}~\bibnamefont {Parcollet}}, \bibinfo {author}
  {\bibfnamefont {D.}~\bibnamefont {Manske}},\ and\ \bibinfo {author}
  {\bibfnamefont {P.}~\bibnamefont {Hansmann}},\ }\bibfield  {title} {\bibinfo
  {title} {Chiral $d$-wave superconductivity in a triangular surface lattice
  mediated by long-range interaction},\ }\href
  {https://doi.org/10.1103/PhysRevB.97.155145} {\bibfield  {journal} {\bibinfo
  {journal} {Phys. Rev. B}\ }\textbf {\bibinfo {volume} {97}},\ \bibinfo
  {pages} {155145} (\bibinfo {year} {2018})}\BibitemShut {NoStop}%
\bibitem [{\citenamefont {Wolf}\ \emph {et~al.}(2022)\citenamefont {Wolf},
  \citenamefont {Di~Sante}, \citenamefont {Schwemmer}, \citenamefont
  {Thomale},\ and\ \citenamefont {Rachel}}]{Wolf2022}%
  \BibitemOpen
  \bibfield  {author} {\bibinfo {author} {\bibfnamefont {S.}~\bibnamefont
  {Wolf}}, \bibinfo {author} {\bibfnamefont {D.}~\bibnamefont {Di~Sante}},
  \bibinfo {author} {\bibfnamefont {T.}~\bibnamefont {Schwemmer}}, \bibinfo
  {author} {\bibfnamefont {R.}~\bibnamefont {Thomale}},\ and\ \bibinfo {author}
  {\bibfnamefont {S.}~\bibnamefont {Rachel}},\ }\bibfield  {title} {\bibinfo
  {title} {Triplet superconductivity from nonlocal {C}oulomb repulsion in an
  atomic {Sn} layer deposited onto a {Si(111)} substrate},\ }\href
  {https://doi.org/10.1103/PhysRevLett.128.167002} {\bibfield  {journal}
  {\bibinfo  {journal} {Phys. Rev. Lett.}\ }\textbf {\bibinfo {volume} {128}},\
  \bibinfo {pages} {167002} (\bibinfo {year} {2022})}\BibitemShut {NoStop}%
\bibitem [{\citenamefont {Zahedifar}\ and\ \citenamefont
  {Kratzer}(2019)}]{ZahedifarPRB2019}%
  \BibitemOpen
  \bibfield  {author} {\bibinfo {author} {\bibfnamefont {M.}~\bibnamefont
  {Zahedifar}}\ and\ \bibinfo {author} {\bibfnamefont {P.}~\bibnamefont
  {Kratzer}},\ }\bibfield  {title} {\bibinfo {title} {Phonon-induced electronic
  relaxation in a strongly correlated system: The {Sn/Si(111)}
  $(\sqrt{3}\ifmmode\times\else\texttimes\fi{}\sqrt{3})$ adlayer revisited},\
  }\href {https://doi.org/10.1103/PhysRevB.100.125427} {\bibfield  {journal}
  {\bibinfo  {journal} {Phys. Rev. B}\ }\textbf {\bibinfo {volume} {100}},\
  \bibinfo {pages} {125427} (\bibinfo {year} {2019})}\BibitemShut {NoStop}%
\bibitem [{\citenamefont {Howald}\ \emph {et~al.}(2003)\citenamefont {Howald},
  \citenamefont {Eisaki}, \citenamefont {Kaneko},\ and\ \citenamefont
  {Kapitulnik}}]{HowaldPNAS2003}%
  \BibitemOpen
  \bibfield  {author} {\bibinfo {author} {\bibfnamefont {C.}~\bibnamefont
  {Howald}}, \bibinfo {author} {\bibfnamefont {H.}~\bibnamefont {Eisaki}},
  \bibinfo {author} {\bibfnamefont {N.}~\bibnamefont {Kaneko}},\ and\ \bibinfo
  {author} {\bibfnamefont {A.}~\bibnamefont {Kapitulnik}},\ }\bibfield  {title}
  {\bibinfo {title} {Coexistence of periodic modulation of quasiparticle states
  and superconductivity in {Bi$_2$Sr$_2$CaCu$_2$O$_{8+\delta}$}},\ }\href
  {https://doi.org/10.1073/pnas.1233768100} {\bibfield  {journal} {\bibinfo
  {journal} {Proceedings of the National Academy of Sciences}\ }\textbf
  {\bibinfo {volume} {100}},\ \bibinfo {pages} {9705} (\bibinfo {year}
  {2003})}\BibitemShut {NoStop}%
\bibitem [{\citenamefont {Vershinin}\ \emph {et~al.}(2004)\citenamefont
  {Vershinin}, \citenamefont {Misra}, \citenamefont {Ono}, \citenamefont {Abe},
  \citenamefont {Ando},\ and\ \citenamefont {Yazdani}}]{Vershinin2004}%
  \BibitemOpen
  \bibfield  {author} {\bibinfo {author} {\bibfnamefont {M.}~\bibnamefont
  {Vershinin}}, \bibinfo {author} {\bibfnamefont {S.}~\bibnamefont {Misra}},
  \bibinfo {author} {\bibfnamefont {S.}~\bibnamefont {Ono}}, \bibinfo {author}
  {\bibfnamefont {Y.}~\bibnamefont {Abe}}, \bibinfo {author} {\bibfnamefont
  {Y.}~\bibnamefont {Ando}},\ and\ \bibinfo {author} {\bibfnamefont
  {A.}~\bibnamefont {Yazdani}},\ }\bibfield  {title} {\bibinfo {title} {Local
  ordering in the pseudogap state of the high-{$T_c$} superconductor
  {Bi$_2$Sr$_2$CaCu$_2$O$_{8+\delta}$}},\ }\href
  {https://doi.org/10.1126/science.1093384} {\bibfield  {journal} {\bibinfo
  {journal} {Science}\ }\textbf {\bibinfo {volume} {303}},\ \bibinfo {pages}
  {1995} (\bibinfo {year} {2004})}\BibitemShut {NoStop}%
\bibitem [{\citenamefont {Dynes}\ \emph {et~al.}(1978)\citenamefont {Dynes},
  \citenamefont {Narayanamurti},\ and\ \citenamefont {Garno}}]{DynesPRL1978}%
  \BibitemOpen
  \bibfield  {author} {\bibinfo {author} {\bibfnamefont {R.~C.}\ \bibnamefont
  {Dynes}}, \bibinfo {author} {\bibfnamefont {V.}~\bibnamefont
  {Narayanamurti}},\ and\ \bibinfo {author} {\bibfnamefont {J.~P.}\
  \bibnamefont {Garno}},\ }\bibfield  {title} {\bibinfo {title} {Direct
  measurement of quasiparticle-lifetime broadening in a strong-coupled
  superconductor},\ }\href {https://doi.org/10.1103/PhysRevLett.41.1509}
  {\bibfield  {journal} {\bibinfo  {journal} {Phys. Rev. Lett.}\ }\textbf
  {\bibinfo {volume} {41}},\ \bibinfo {pages} {1509} (\bibinfo {year}
  {1978})}\BibitemShut {NoStop}%
\bibitem [{\citenamefont {Petersen}\ \emph {et~al.}(2000)\citenamefont
  {Petersen}, \citenamefont {Hofmann}, \citenamefont {Plummer},\ and\
  \citenamefont {Besenbacher}}]{Petersen2002JESRP}%
  \BibitemOpen
  \bibfield  {author} {\bibinfo {author} {\bibfnamefont {L.}~\bibnamefont
  {Petersen}}, \bibinfo {author} {\bibfnamefont {P.}~\bibnamefont {Hofmann}},
  \bibinfo {author} {\bibfnamefont {E.}~\bibnamefont {Plummer}},\ and\ \bibinfo
  {author} {\bibfnamefont {F.}~\bibnamefont {Besenbacher}},\ }\bibfield
  {title} {\bibinfo {title} {Fourier transform--{STM}: determining the surface
  {F}ermi contour},\ }\href
  {https://doi.org/https://doi.org/10.1016/S0368-2048(00)00110-9} {\bibfield
  {journal} {\bibinfo  {journal} {Journal of Electron Spectroscopy and Related
  Phenomena}\ }\textbf {\bibinfo {volume} {109}},\ \bibinfo {pages} {97}
  (\bibinfo {year} {2000})}\BibitemShut {NoStop}%
\bibitem [{\citenamefont {Yu}(1965)}]{Yu}%
  \BibitemOpen
  \bibfield  {author} {\bibinfo {author} {\bibfnamefont {L.}~\bibnamefont
  {Yu}},\ }\bibfield  {title} {\bibinfo {title} {Bound state in superconductors
  with paramagnetic impurities},\ }\href {https://www.osti.gov/biblio/4638963}
  {\bibfield  {journal} {\bibinfo  {journal} {Wu Li Hsueh Pao (China)
  Supersedes Chung-Kuo Wu Li Hsueh For English translation see Chin. J. Phys.
  (Peking) (Engl. Transl.)}\ }\textbf {\bibinfo {volume} {21}} (\bibinfo {year}
  {1965})}\BibitemShut {NoStop}%
\bibitem [{\citenamefont {Shiba}(1968)}]{Shiba}%
  \BibitemOpen
  \bibfield  {author} {\bibinfo {author} {\bibfnamefont {H.}~\bibnamefont
  {Shiba}},\ }\bibfield  {title} {\bibinfo {title} {{Classical Spins in
  Superconductors}},\ }\href {https://doi.org/10.1143/PTP.40.435} {\bibfield
  {journal} {\bibinfo  {journal} {Progress of Theoretical Physics}\ }\textbf
  {\bibinfo {volume} {40}},\ \bibinfo {pages} {435} (\bibinfo {year}
  {1968})}\BibitemShut {NoStop}%
\bibitem [{\citenamefont {Rusinov}(1969)}]{Rusinov}%
  \BibitemOpen
  \bibfield  {author} {\bibinfo {author} {\bibfnamefont {A.~I.}\ \bibnamefont
  {Rusinov}},\ }\bibfield  {title} {\bibinfo {title} {Superconductivity near a
  paramagnetic impurity},\ }\href {https://www.osti.gov/biblio/4780988}
  {\bibfield  {journal} {\bibinfo  {journal} {JETP Lett. (USSR) (Engl.
  Transl.), 9: 85-7 (Jan. 20, 1969).}\ } (\bibinfo {year} {1969})}\BibitemShut
  {NoStop}%
\bibitem [{\citenamefont {M{\'e}nard}\ \emph {et~al.}(2015)\citenamefont
  {M{\'e}nard}, \citenamefont {Guissart}, \citenamefont {Brun}, \citenamefont
  {Pons}, \citenamefont {Stolyarov}, \citenamefont {Debontridder},
  \citenamefont {Leclerc}, \citenamefont {Janod}, \citenamefont {Cario},
  \citenamefont {Roditchev}, \citenamefont {Simon},\ and\ \citenamefont
  {Cren}}]{MenardNP2015}%
  \BibitemOpen
  \bibfield  {author} {\bibinfo {author} {\bibfnamefont {G.~C.}\ \bibnamefont
  {M{\'e}nard}}, \bibinfo {author} {\bibfnamefont {S.}~\bibnamefont
  {Guissart}}, \bibinfo {author} {\bibfnamefont {C.}~\bibnamefont {Brun}},
  \bibinfo {author} {\bibfnamefont {S.}~\bibnamefont {Pons}}, \bibinfo {author}
  {\bibfnamefont {V.~S.}\ \bibnamefont {Stolyarov}}, \bibinfo {author}
  {\bibfnamefont {F.}~\bibnamefont {Debontridder}}, \bibinfo {author}
  {\bibfnamefont {M.~V.}\ \bibnamefont {Leclerc}}, \bibinfo {author}
  {\bibfnamefont {E.}~\bibnamefont {Janod}}, \bibinfo {author} {\bibfnamefont
  {L.}~\bibnamefont {Cario}}, \bibinfo {author} {\bibfnamefont
  {D.}~\bibnamefont {Roditchev}}, \bibinfo {author} {\bibfnamefont
  {P.}~\bibnamefont {Simon}},\ and\ \bibinfo {author} {\bibfnamefont
  {T.}~\bibnamefont {Cren}},\ }\bibfield  {title} {\bibinfo {title} {Coherent
  long-range magnetic bound states in a superconductor},\ }\href
  {https://doi.org/10.1038/nphys3508} {\bibfield  {journal} {\bibinfo
  {journal} {Nature Physics}\ }\textbf {\bibinfo {volume} {11}},\ \bibinfo
  {pages} {1013} (\bibinfo {year} {2015})}\BibitemShut {NoStop}%
\bibitem [{\citenamefont {Kim}\ \emph {et~al.}(2020)\citenamefont {Kim},
  \citenamefont {R{\'o}zsa}, \citenamefont {Schreyer}, \citenamefont {Simon},\
  and\ \citenamefont {Wiesendanger}}]{KimNC2020}%
  \BibitemOpen
  \bibfield  {author} {\bibinfo {author} {\bibfnamefont {H.}~\bibnamefont
  {Kim}}, \bibinfo {author} {\bibfnamefont {L.}~\bibnamefont {R{\'o}zsa}},
  \bibinfo {author} {\bibfnamefont {D.}~\bibnamefont {Schreyer}}, \bibinfo
  {author} {\bibfnamefont {E.}~\bibnamefont {Simon}},\ and\ \bibinfo {author}
  {\bibfnamefont {R.}~\bibnamefont {Wiesendanger}},\ }\bibfield  {title}
  {\bibinfo {title} {Long-range focusing of magnetic bound states in
  superconducting lanthanum},\ }\href
  {https://doi.org/10.1038/s41467-020-18406-8} {\bibfield  {journal} {\bibinfo
  {journal} {Nature Communications}\ }\textbf {\bibinfo {volume} {11}},\
  \bibinfo {pages} {4573} (\bibinfo {year} {2020})}\BibitemShut {NoStop}%
\bibitem [{\citenamefont {Wang}\ and\ \citenamefont
  {Wang}(2004)}]{WangPRB2004}%
  \BibitemOpen
  \bibfield  {author} {\bibinfo {author} {\bibfnamefont {Q.-H.}\ \bibnamefont
  {Wang}}\ and\ \bibinfo {author} {\bibfnamefont {Z.~D.}\ \bibnamefont
  {Wang}},\ }\bibfield  {title} {\bibinfo {title} {Impurity and interface bound
  states in ${d}_{{x}^{2}\ensuremath{-}{y}^{2}}{+id}_{\mathrm{xy}}$ and
  ${p}_{x}{+ip}_{y}$ superconductors},\ }\href
  {https://doi.org/10.1103/PhysRevB.69.092502} {\bibfield  {journal} {\bibinfo
  {journal} {Phys. Rev. B}\ }\textbf {\bibinfo {volume} {69}},\ \bibinfo
  {pages} {092502} (\bibinfo {year} {2004})}\BibitemShut {NoStop}%
\bibitem [{\citenamefont {Mashkoori}\ \emph {et~al.}(2017)\citenamefont
  {Mashkoori}, \citenamefont {Bj{\"o}rnson},\ and\ \citenamefont
  {Black-Schaffer}}]{MashkooriSSR2017}%
  \BibitemOpen
  \bibfield  {author} {\bibinfo {author} {\bibfnamefont {M.}~\bibnamefont
  {Mashkoori}}, \bibinfo {author} {\bibfnamefont {K.}~\bibnamefont
  {Bj{\"o}rnson}},\ and\ \bibinfo {author} {\bibfnamefont {A.~M.}\ \bibnamefont
  {Black-Schaffer}},\ }\bibfield  {title} {\bibinfo {title} {Impurity bound
  states in fully gapped $d$-wave superconductors with subdominant order
  parameters},\ }\href {https://doi.org/10.1038/srep44107} {\bibfield
  {journal} {\bibinfo  {journal} {Scientific Reports}\ }\textbf {\bibinfo
  {volume} {7}},\ \bibinfo {pages} {44107} (\bibinfo {year}
  {2017})}\BibitemShut {NoStop}%
\bibitem [{\citenamefont {Anderson}(1959)}]{Anderson}%
  \BibitemOpen
  \bibfield  {author} {\bibinfo {author} {\bibfnamefont {P.}~\bibnamefont
  {Anderson}},\ }\bibfield  {title} {\bibinfo {title} {Theory of dirty
  superconductors},\ }\href
  {https://doi.org/https://doi.org/10.1016/0022-3697(59)90036-8} {\bibfield
  {journal} {\bibinfo  {journal} {Journal of Physics and Chemistry of Solids}\
  }\textbf {\bibinfo {volume} {11}},\ \bibinfo {pages} {26} (\bibinfo {year}
  {1959})}\BibitemShut {NoStop}%
\bibitem [{\citenamefont {Maier}\ \emph {et~al.}(2006)\citenamefont {Maier},
  \citenamefont {Jarrell},\ and\ \citenamefont {Scalapino}}]{MaierPRL2006}%
  \BibitemOpen
  \bibfield  {author} {\bibinfo {author} {\bibfnamefont {T.~A.}\ \bibnamefont
  {Maier}}, \bibinfo {author} {\bibfnamefont {M.~S.}\ \bibnamefont {Jarrell}},\
  and\ \bibinfo {author} {\bibfnamefont {D.~J.}\ \bibnamefont {Scalapino}},\
  }\bibfield  {title} {\bibinfo {title} {Structure of the pairing interaction
  in the two-dimensional {H}ubbard model},\ }\href
  {https://doi.org/10.1103/PhysRevLett.96.047005} {\bibfield  {journal}
  {\bibinfo  {journal} {Phys. Rev. Lett.}\ }\textbf {\bibinfo {volume} {96}},\
  \bibinfo {pages} {047005} (\bibinfo {year} {2006})}\BibitemShut {NoStop}%
\bibitem [{\citenamefont {H{\"a}hner}\ \emph {et~al.}(2020)\citenamefont
  {H{\"a}hner}, \citenamefont {Alvarez}, \citenamefont {Maier}, \citenamefont
  {Solc{\`a}}, \citenamefont {Staar}, \citenamefont {Summers},\ and\
  \citenamefont {Schulthess}}]{UrsCompPhysComm2020}%
  \BibitemOpen
  \bibfield  {author} {\bibinfo {author} {\bibfnamefont {U.~R.}\ \bibnamefont
  {H{\"a}hner}}, \bibinfo {author} {\bibfnamefont {G.}~\bibnamefont {Alvarez}},
  \bibinfo {author} {\bibfnamefont {T.~A.}\ \bibnamefont {Maier}}, \bibinfo
  {author} {\bibfnamefont {R.}~\bibnamefont {Solc{\`a}}}, \bibinfo {author}
  {\bibfnamefont {P.}~\bibnamefont {Staar}}, \bibinfo {author} {\bibfnamefont
  {M.~S.}\ \bibnamefont {Summers}},\ and\ \bibinfo {author} {\bibfnamefont
  {T.~C.}\ \bibnamefont {Schulthess}},\ }\bibfield  {title} {\bibinfo {title}
  {{DCA++}: A software framework to solve correlated electron problems with
  modern quantum cluster methods},\ }\href
  {https://doi.org/https://doi.org/10.1016/j.cpc.2019.01.006} {\bibfield
  {journal} {\bibinfo  {journal} {Computer Physics Communications}\ }\textbf
  {\bibinfo {volume} {246}},\ \bibinfo {pages} {106709} (\bibinfo {year}
  {2020})}\BibitemShut {NoStop}%
\bibitem [{\citenamefont {Chen}\ \emph {et~al.}(2013)\citenamefont {Chen},
  \citenamefont {Meng}, \citenamefont {Yu}, \citenamefont {Yang}, \citenamefont
  {Jarrell},\ and\ \citenamefont {Moreno}}]{ChenPRB2013}%
  \BibitemOpen
  \bibfield  {author} {\bibinfo {author} {\bibfnamefont {K.~S.}\ \bibnamefont
  {Chen}}, \bibinfo {author} {\bibfnamefont {Z.~Y.}\ \bibnamefont {Meng}},
  \bibinfo {author} {\bibfnamefont {U.}~\bibnamefont {Yu}}, \bibinfo {author}
  {\bibfnamefont {S.}~\bibnamefont {Yang}}, \bibinfo {author} {\bibfnamefont
  {M.}~\bibnamefont {Jarrell}},\ and\ \bibinfo {author} {\bibfnamefont
  {J.}~\bibnamefont {Moreno}},\ }\bibfield  {title} {\bibinfo {title}
  {Unconventional superconductivity on the triangular lattice {H}ubbard
  model},\ }\href {https://doi.org/10.1103/PhysRevB.88.041103} {\bibfield
  {journal} {\bibinfo  {journal} {Phys. Rev. B}\ }\textbf {\bibinfo {volume}
  {88}},\ \bibinfo {pages} {041103} (\bibinfo {year} {2013})}\BibitemShut
  {NoStop}%
\bibitem [{\citenamefont {Volovik}(1997)}]{VolovikJETP1997}%
  \BibitemOpen
  \bibfield  {author} {\bibinfo {author} {\bibfnamefont {G.~E.}\ \bibnamefont
  {Volovik}},\ }\bibfield  {title} {\bibinfo {title} {On edge states in
  superconductors with time inversion symmetry breaking},\ }\href
  {https://doi.org/10.1134/1.567563} {\bibfield  {journal} {\bibinfo  {journal}
  {Journal of Experimental and Theoretical Physics Letters}\ }\textbf {\bibinfo
  {volume} {66}},\ \bibinfo {pages} {522} (\bibinfo {year} {1997})}\BibitemShut
  {NoStop}%
\bibitem [{\citenamefont {Laughlin}(1998)}]{LaughlinPRL1998}%
  \BibitemOpen
  \bibfield  {author} {\bibinfo {author} {\bibfnamefont {R.~B.}\ \bibnamefont
  {Laughlin}},\ }\bibfield  {title} {\bibinfo {title} {Magnetic induction of
  ${\mathit{d}}_{{\mathit{x}}^{2}\ensuremath{-}{\mathit{y}}^{2}}+{\mathrm{id}}_{\mathrm{xy}}$
  order in high-${T}_{c}$ superconductors},\ }\href
  {https://doi.org/10.1103/PhysRevLett.80.5188} {\bibfield  {journal} {\bibinfo
   {journal} {Phys. Rev. Lett.}\ }\textbf {\bibinfo {volume} {80}},\ \bibinfo
  {pages} {5188} (\bibinfo {year} {1998})}\BibitemShut {NoStop}%
\bibitem [{\citenamefont {Senthil}\ \emph {et~al.}(1999)\citenamefont
  {Senthil}, \citenamefont {Marston},\ and\ \citenamefont
  {Fisher}}]{SenthilPRB1999}%
  \BibitemOpen
  \bibfield  {author} {\bibinfo {author} {\bibfnamefont {T.}~\bibnamefont
  {Senthil}}, \bibinfo {author} {\bibfnamefont {J.~B.}\ \bibnamefont
  {Marston}},\ and\ \bibinfo {author} {\bibfnamefont {M.~P.~A.}\ \bibnamefont
  {Fisher}},\ }\bibfield  {title} {\bibinfo {title} {Spin quantum {H}all effect
  in unconventional superconductors},\ }\href
  {https://doi.org/10.1103/PhysRevB.60.4245} {\bibfield  {journal} {\bibinfo
  {journal} {Phys. Rev. B}\ }\textbf {\bibinfo {volume} {60}},\ \bibinfo
  {pages} {4245} (\bibinfo {year} {1999})}\BibitemShut {NoStop}%
\bibitem [{\citenamefont {Ming}\ \emph
  {et~al.}(2017{\natexlab{b}})\citenamefont {Ming}, \citenamefont {Mulugeta},
  \citenamefont {Tu}, \citenamefont {Smith}, \citenamefont {Vilmercati},
  \citenamefont {Lee}, \citenamefont {Huang}, \citenamefont {Diehl},
  \citenamefont {Snijders},\ and\ \citenamefont {Weitering}}]{MingNC2017}%
  \BibitemOpen
  \bibfield  {author} {\bibinfo {author} {\bibfnamefont {F.}~\bibnamefont
  {Ming}}, \bibinfo {author} {\bibfnamefont {D.}~\bibnamefont {Mulugeta}},
  \bibinfo {author} {\bibfnamefont {W.}~\bibnamefont {Tu}}, \bibinfo {author}
  {\bibfnamefont {T.~S.}\ \bibnamefont {Smith}}, \bibinfo {author}
  {\bibfnamefont {P.}~\bibnamefont {Vilmercati}}, \bibinfo {author}
  {\bibfnamefont {G.}~\bibnamefont {Lee}}, \bibinfo {author} {\bibfnamefont
  {Y.-T.}\ \bibnamefont {Huang}}, \bibinfo {author} {\bibfnamefont {R.~D.}\
  \bibnamefont {Diehl}}, \bibinfo {author} {\bibfnamefont {P.~C.}\ \bibnamefont
  {Snijders}},\ and\ \bibinfo {author} {\bibfnamefont {H.~H.}\ \bibnamefont
  {Weitering}},\ }\bibfield  {title} {\bibinfo {title} {Hidden phase in a
  two-dimensional {Sn} layer stabilized by modulation hole doping},\ }\href
  {https://doi.org/10.1038/ncomms14721} {\bibfield  {journal} {\bibinfo
  {journal} {Nature Communications}\ }\textbf {\bibinfo {volume} {8}},\
  \bibinfo {pages} {14721} (\bibinfo {year} {2017}{\natexlab{b}})}\BibitemShut
  {NoStop}%
\bibitem [{\citenamefont {De~Gennes}(1964)}]{DeGennesRMP1964}%
  \BibitemOpen
  \bibfield  {author} {\bibinfo {author} {\bibfnamefont {P.~G.}\ \bibnamefont
  {De~Gennes}},\ }\bibfield  {title} {\bibinfo {title} {Boundary effects in
  superconductors},\ }\href {https://doi.org/10.1103/RevModPhys.36.225}
  {\bibfield  {journal} {\bibinfo  {journal} {Rev. Mod. Phys.}\ }\textbf
  {\bibinfo {volume} {36}},\ \bibinfo {pages} {225} (\bibinfo {year}
  {1964})}\BibitemShut {NoStop}%
\bibitem [{\citenamefont {Zhou}\ and\ \citenamefont
  {Wang}(2008)}]{ZhouPRL2008}%
  \BibitemOpen
  \bibfield  {author} {\bibinfo {author} {\bibfnamefont {S.}~\bibnamefont
  {Zhou}}\ and\ \bibinfo {author} {\bibfnamefont {Z.}~\bibnamefont {Wang}},\
  }\bibfield  {title} {\bibinfo {title} {Nodal $d+id$ pairing and topological
  phases on the triangular lattice of
  {${\mathrm{Na}}_{x}{\mathrm{CoO}}_{2}\ifmmode\cdot\else\textperiodcentered\fi{}y{\mathrm{H}}_{2}\mathrm{O}$}:
  Evidence for an unconventional superconducting state},\ }\href
  {https://doi.org/10.1103/PhysRevLett.100.217002} {\bibfield  {journal}
  {\bibinfo  {journal} {Phys. Rev. Lett.}\ }\textbf {\bibinfo {volume} {100}},\
  \bibinfo {pages} {217002} (\bibinfo {year} {2008})}\BibitemShut {NoStop}%
\bibitem [{\citenamefont {Giannozzi}\ \emph {et~al.}(2009)\citenamefont
  {Giannozzi}, \citenamefont {Baroni}, \citenamefont {Bonini}, \citenamefont
  {Calandra}, \citenamefont {Car}, \citenamefont {Cavazzoni}, \citenamefont
  {Ceresoli}, \citenamefont {Chiarotti}, \citenamefont {Cococcioni},
  \citenamefont {Dabo}, \citenamefont {Corso}, \citenamefont {de~Gironcoli},
  \citenamefont {Fabris}, \citenamefont {Fratesi}, \citenamefont {Gebauer},
  \citenamefont {Gerstmann}, \citenamefont {Gougoussis}, \citenamefont
  {Kokalj}, \citenamefont {Lazzeri}, \citenamefont {Martin-Samos},
  \citenamefont {Marzari}, \citenamefont {Mauri}, \citenamefont {Mazzarello},
  \citenamefont {Paolini}, \citenamefont {Pasquarello}, \citenamefont
  {Paulatto}, \citenamefont {Sbraccia}, \citenamefont {Scandolo}, \citenamefont
  {Sclauzero}, \citenamefont {Seitsonen}, \citenamefont {Smogunov},
  \citenamefont {Umari},\ and\ \citenamefont
  {Wentzcovitch}}]{GiannozziJPCM2009}%
  \BibitemOpen
  \bibfield  {author} {\bibinfo {author} {\bibfnamefont {P.}~\bibnamefont
  {Giannozzi}}, \bibinfo {author} {\bibfnamefont {S.}~\bibnamefont {Baroni}},
  \bibinfo {author} {\bibfnamefont {N.}~\bibnamefont {Bonini}}, \bibinfo
  {author} {\bibfnamefont {M.}~\bibnamefont {Calandra}}, \bibinfo {author}
  {\bibfnamefont {R.}~\bibnamefont {Car}}, \bibinfo {author} {\bibfnamefont
  {C.}~\bibnamefont {Cavazzoni}}, \bibinfo {author} {\bibfnamefont
  {D.}~\bibnamefont {Ceresoli}}, \bibinfo {author} {\bibfnamefont {G.~L.}\
  \bibnamefont {Chiarotti}}, \bibinfo {author} {\bibfnamefont {M.}~\bibnamefont
  {Cococcioni}}, \bibinfo {author} {\bibfnamefont {I.}~\bibnamefont {Dabo}},
  \bibinfo {author} {\bibfnamefont {A.~D.}\ \bibnamefont {Corso}}, \bibinfo
  {author} {\bibfnamefont {S.}~\bibnamefont {de~Gironcoli}}, \bibinfo {author}
  {\bibfnamefont {S.}~\bibnamefont {Fabris}}, \bibinfo {author} {\bibfnamefont
  {G.}~\bibnamefont {Fratesi}}, \bibinfo {author} {\bibfnamefont
  {R.}~\bibnamefont {Gebauer}}, \bibinfo {author} {\bibfnamefont
  {U.}~\bibnamefont {Gerstmann}}, \bibinfo {author} {\bibfnamefont
  {C.}~\bibnamefont {Gougoussis}}, \bibinfo {author} {\bibfnamefont
  {A.}~\bibnamefont {Kokalj}}, \bibinfo {author} {\bibfnamefont
  {M.}~\bibnamefont {Lazzeri}}, \bibinfo {author} {\bibfnamefont
  {L.}~\bibnamefont {Martin-Samos}}, \bibinfo {author} {\bibfnamefont
  {N.}~\bibnamefont {Marzari}}, \bibinfo {author} {\bibfnamefont
  {F.}~\bibnamefont {Mauri}}, \bibinfo {author} {\bibfnamefont
  {R.}~\bibnamefont {Mazzarello}}, \bibinfo {author} {\bibfnamefont
  {S.}~\bibnamefont {Paolini}}, \bibinfo {author} {\bibfnamefont
  {A.}~\bibnamefont {Pasquarello}}, \bibinfo {author} {\bibfnamefont
  {L.}~\bibnamefont {Paulatto}}, \bibinfo {author} {\bibfnamefont
  {C.}~\bibnamefont {Sbraccia}}, \bibinfo {author} {\bibfnamefont
  {S.}~\bibnamefont {Scandolo}}, \bibinfo {author} {\bibfnamefont
  {G.}~\bibnamefont {Sclauzero}}, \bibinfo {author} {\bibfnamefont {A.~P.}\
  \bibnamefont {Seitsonen}}, \bibinfo {author} {\bibfnamefont {A.}~\bibnamefont
  {Smogunov}}, \bibinfo {author} {\bibfnamefont {P.}~\bibnamefont {Umari}},\
  and\ \bibinfo {author} {\bibfnamefont {R.~M.}\ \bibnamefont {Wentzcovitch}},\
  }\bibfield  {title} {\bibinfo {title} {{QUANTUM} {ESPRESSO}: a modular and
  open-source software project for quantum simulations of materials},\ }\href
  {https://doi.org/10.1088/0953-8984/21/39/395502} {\bibfield  {journal}
  {\bibinfo  {journal} {Journal of Physics: Condensed Matter}\ }\textbf
  {\bibinfo {volume} {21}},\ \bibinfo {pages} {395502} (\bibinfo {year}
  {2009})}\BibitemShut {NoStop}%
\bibitem [{\citenamefont {Perdew}\ \emph {et~al.}(1996)\citenamefont {Perdew},
  \citenamefont {Burke},\ and\ \citenamefont {Ernzerhof}}]{PerdewPRL1996}%
  \BibitemOpen
  \bibfield  {author} {\bibinfo {author} {\bibfnamefont {J.~P.}\ \bibnamefont
  {Perdew}}, \bibinfo {author} {\bibfnamefont {K.}~\bibnamefont {Burke}},\ and\
  \bibinfo {author} {\bibfnamefont {M.}~\bibnamefont {Ernzerhof}},\ }\bibfield
  {title} {\bibinfo {title} {Generalized gradient approximation made simple},\
  }\href {https://doi.org/10.1103/PhysRevLett.77.3865} {\bibfield  {journal}
  {\bibinfo  {journal} {Phys. Rev. Lett.}\ }\textbf {\bibinfo {volume} {77}},\
  \bibinfo {pages} {3865} (\bibinfo {year} {1996})}\BibitemShut {NoStop}%
\bibitem [{\citenamefont {Goedecker}\ \emph {et~al.}(1996)\citenamefont
  {Goedecker}, \citenamefont {Teter},\ and\ \citenamefont
  {Hutter}}]{GoedeckerPRB1996}%
  \BibitemOpen
  \bibfield  {author} {\bibinfo {author} {\bibfnamefont {S.}~\bibnamefont
  {Goedecker}}, \bibinfo {author} {\bibfnamefont {M.}~\bibnamefont {Teter}},\
  and\ \bibinfo {author} {\bibfnamefont {J.}~\bibnamefont {Hutter}},\
  }\bibfield  {title} {\bibinfo {title} {Separable dual-space gaussian
  pseudopotentials},\ }\href {https://doi.org/10.1103/PhysRevB.54.1703}
  {\bibfield  {journal} {\bibinfo  {journal} {Phys. Rev. B}\ }\textbf {\bibinfo
  {volume} {54}},\ \bibinfo {pages} {1703} (\bibinfo {year}
  {1996})}\BibitemShut {NoStop}%
\bibitem [{\citenamefont {Hartwigsen}\ \emph {et~al.}(1998)\citenamefont
  {Hartwigsen}, \citenamefont {Goedecker},\ and\ \citenamefont
  {Hutter}}]{HartwigsenPRB1998}%
  \BibitemOpen
  \bibfield  {author} {\bibinfo {author} {\bibfnamefont {C.}~\bibnamefont
  {Hartwigsen}}, \bibinfo {author} {\bibfnamefont {S.}~\bibnamefont
  {Goedecker}},\ and\ \bibinfo {author} {\bibfnamefont {J.}~\bibnamefont
  {Hutter}},\ }\bibfield  {title} {\bibinfo {title} {Relativistic separable
  dual-space gaussian pseudopotentials from {H} to {Rn}},\ }\href
  {https://doi.org/10.1103/PhysRevB.58.3641} {\bibfield  {journal} {\bibinfo
  {journal} {Phys. Rev. B}\ }\textbf {\bibinfo {volume} {58}},\ \bibinfo
  {pages} {3641} (\bibinfo {year} {1998})}\BibitemShut {NoStop}%
\bibitem [{\citenamefont {P\'erez}\ \emph {et~al.}(2001)\citenamefont
  {P\'erez}, \citenamefont {Ortega},\ and\ \citenamefont
  {Flores}}]{PerezPRL2001}%
  \BibitemOpen
  \bibfield  {author} {\bibinfo {author} {\bibfnamefont {R.}~\bibnamefont
  {P\'erez}}, \bibinfo {author} {\bibfnamefont {J.}~\bibnamefont {Ortega}},\
  and\ \bibinfo {author} {\bibfnamefont {F.}~\bibnamefont {Flores}},\
  }\bibfield  {title} {\bibinfo {title} {Surface soft phonon and the
  $\ensuremath{\surd}3\ifmmode\times\else\texttimes\fi{}\ensuremath{\surd}3\ensuremath{\leftrightarrow}3\ifmmode\times\else\texttimes\fi{}3$
  phase transition in $\mathrm{Sn}/\mathrm{Ge}(111)$ and
  $\mathrm{Sn}/\mathrm{Si}(111)$},\ }\href
  {https://doi.org/10.1103/PhysRevLett.86.4891} {\bibfield  {journal} {\bibinfo
   {journal} {Phys. Rev. Lett.}\ }\textbf {\bibinfo {volume} {86}},\ \bibinfo
  {pages} {4891} (\bibinfo {year} {2001})}\BibitemShut {NoStop}%
\bibitem [{\citenamefont {Monkhorst}\ and\ \citenamefont
  {Pack}(1976)}]{MonkhorstPRB1976}%
  \BibitemOpen
  \bibfield  {author} {\bibinfo {author} {\bibfnamefont {H.~J.}\ \bibnamefont
  {Monkhorst}}\ and\ \bibinfo {author} {\bibfnamefont {J.~D.}\ \bibnamefont
  {Pack}},\ }\bibfield  {title} {\bibinfo {title} {Special points for
  {B}rillouin-zone integrations},\ }\href
  {https://doi.org/10.1103/PhysRevB.13.5188} {\bibfield  {journal} {\bibinfo
  {journal} {Phys. Rev. B}\ }\textbf {\bibinfo {volume} {13}},\ \bibinfo
  {pages} {5188} (\bibinfo {year} {1976})}\BibitemShut {NoStop}%
\bibitem [{\citenamefont {Blanco}\ \emph {et~al.}(2004)\citenamefont {Blanco},
  \citenamefont {Gonz\'alez}, \citenamefont {Jel\'{\i}nek}, \citenamefont
  {Ortega}, \citenamefont {Flores},\ and\ \citenamefont
  {P\'erez}}]{BlancoPRB2004}%
  \BibitemOpen
  \bibfield  {author} {\bibinfo {author} {\bibfnamefont {J.~M.}\ \bibnamefont
  {Blanco}}, \bibinfo {author} {\bibfnamefont {C.}~\bibnamefont {Gonz\'alez}},
  \bibinfo {author} {\bibfnamefont {P.}~\bibnamefont {Jel\'{\i}nek}}, \bibinfo
  {author} {\bibfnamefont {J.}~\bibnamefont {Ortega}}, \bibinfo {author}
  {\bibfnamefont {F.}~\bibnamefont {Flores}},\ and\ \bibinfo {author}
  {\bibfnamefont {R.}~\bibnamefont {P\'erez}},\ }\bibfield  {title} {\bibinfo
  {title} {First-principles simulations of {STM} images: From tunneling to the
  contact regime},\ }\href {https://doi.org/10.1103/PhysRevB.70.085405}
  {\bibfield  {journal} {\bibinfo  {journal} {Phys. Rev. B}\ }\textbf {\bibinfo
  {volume} {70}},\ \bibinfo {pages} {085405} (\bibinfo {year}
  {2004})}\BibitemShut {NoStop}%
\bibitem [{\citenamefont {Lewis}\ \emph {et~al.}(2011)\citenamefont {Lewis},
  \citenamefont {Jelínek}, \citenamefont {Ortega}, \citenamefont {Demkov},
  \citenamefont {Trabada}, \citenamefont {Haycock}, \citenamefont {Wang},
  \citenamefont {Adams}, \citenamefont {Tomfohr}, \citenamefont {Abad},
  \citenamefont {Wang},\ and\ \citenamefont {Drabold}}]{LewisPSSB2011}%
  \BibitemOpen
  \bibfield  {author} {\bibinfo {author} {\bibfnamefont {J.~P.}\ \bibnamefont
  {Lewis}}, \bibinfo {author} {\bibfnamefont {P.}~\bibnamefont {Jelínek}},
  \bibinfo {author} {\bibfnamefont {J.}~\bibnamefont {Ortega}}, \bibinfo
  {author} {\bibfnamefont {A.~A.}\ \bibnamefont {Demkov}}, \bibinfo {author}
  {\bibfnamefont {D.~G.}\ \bibnamefont {Trabada}}, \bibinfo {author}
  {\bibfnamefont {B.}~\bibnamefont {Haycock}}, \bibinfo {author} {\bibfnamefont
  {H.}~\bibnamefont {Wang}}, \bibinfo {author} {\bibfnamefont {G.}~\bibnamefont
  {Adams}}, \bibinfo {author} {\bibfnamefont {J.~K.}\ \bibnamefont {Tomfohr}},
  \bibinfo {author} {\bibfnamefont {E.}~\bibnamefont {Abad}}, \bibinfo {author}
  {\bibfnamefont {H.}~\bibnamefont {Wang}},\ and\ \bibinfo {author}
  {\bibfnamefont {D.~A.}\ \bibnamefont {Drabold}},\ }\bibfield  {title}
  {\bibinfo {title} {Advances and applications in the {FIREBALL} ab initio
  tight-binding molecular-dynamics formalism},\ }\href
  {https://doi.org/https://doi.org/10.1002/pssb.201147259} {\bibfield
  {journal} {\bibinfo  {journal} {physica status solidi (b)}\ }\textbf
  {\bibinfo {volume} {248}},\ \bibinfo {pages} {1989} (\bibinfo {year}
  {2011})}\BibitemShut {NoStop}%
\bibitem [{\citenamefont {Gonz\'alez}\ \emph {et~al.}(2004)\citenamefont
  {Gonz\'alez}, \citenamefont {Snijders}, \citenamefont {Ortega}, \citenamefont
  {P\'erez}, \citenamefont {Flores}, \citenamefont {Rogge},\ and\ \citenamefont
  {Weitering}}]{GonzalezPRL2004}%
  \BibitemOpen
  \bibfield  {author} {\bibinfo {author} {\bibfnamefont {C.}~\bibnamefont
  {Gonz\'alez}}, \bibinfo {author} {\bibfnamefont {P.~C.}\ \bibnamefont
  {Snijders}}, \bibinfo {author} {\bibfnamefont {J.}~\bibnamefont {Ortega}},
  \bibinfo {author} {\bibfnamefont {R.}~\bibnamefont {P\'erez}}, \bibinfo
  {author} {\bibfnamefont {F.}~\bibnamefont {Flores}}, \bibinfo {author}
  {\bibfnamefont {S.}~\bibnamefont {Rogge}},\ and\ \bibinfo {author}
  {\bibfnamefont {H.~H.}\ \bibnamefont {Weitering}},\ }\bibfield  {title}
  {\bibinfo {title} {Formation of atom wires on vicinal silicon},\ }\href
  {https://doi.org/10.1103/PhysRevLett.93.126106} {\bibfield  {journal}
  {\bibinfo  {journal} {Phys. Rev. Lett.}\ }\textbf {\bibinfo {volume} {93}},\
  \bibinfo {pages} {126106} (\bibinfo {year} {2004})}\BibitemShut {NoStop}%
\bibitem [{\citenamefont {Ming}\ \emph {et~al.}(2018)\citenamefont {Ming},
  \citenamefont {Smith}, \citenamefont {Johnston}, \citenamefont {Snijders},\
  and\ \citenamefont {Weitering}}]{MingPRB2018}%
  \BibitemOpen
  \bibfield  {author} {\bibinfo {author} {\bibfnamefont {F.}~\bibnamefont
  {Ming}}, \bibinfo {author} {\bibfnamefont {T.~S.}\ \bibnamefont {Smith}},
  \bibinfo {author} {\bibfnamefont {S.}~\bibnamefont {Johnston}}, \bibinfo
  {author} {\bibfnamefont {P.~C.}\ \bibnamefont {Snijders}},\ and\ \bibinfo
  {author} {\bibfnamefont {H.~H.}\ \bibnamefont {Weitering}},\ }\bibfield
  {title} {\bibinfo {title} {Zero-bias anomaly in nanoscale hole-doped {M}ott
  insulators on a triangular silicon surface},\ }\href
  {https://doi.org/10.1103/PhysRevB.97.075403} {\bibfield  {journal} {\bibinfo
  {journal} {Phys. Rev. B}\ }\textbf {\bibinfo {volume} {97}},\ \bibinfo
  {pages} {075403} (\bibinfo {year} {2018})}\BibitemShut {NoStop}%
\bibitem [{\citenamefont {Headrick}\ \emph {et~al.}(1989)\citenamefont
  {Headrick}, \citenamefont {Robinson}, \citenamefont {Vlieg},\ and\
  \citenamefont {Feldman}}]{HeadrickPRL1989}%
  \BibitemOpen
  \bibfield  {author} {\bibinfo {author} {\bibfnamefont {R.~L.}\ \bibnamefont
  {Headrick}}, \bibinfo {author} {\bibfnamefont {I.~K.}\ \bibnamefont
  {Robinson}}, \bibinfo {author} {\bibfnamefont {E.}~\bibnamefont {Vlieg}},\
  and\ \bibinfo {author} {\bibfnamefont {L.~C.}\ \bibnamefont {Feldman}},\
  }\bibfield  {title} {\bibinfo {title} {Structure determination of the
  {Si(111):B(\ensuremath{\surd}3\ifmmode\times\else\texttimes\fi{}\ensuremath{\surd}3)R30\ifmmode^\circ\else\textdegree\fi{}}
  surface: Subsurface substitutional doping},\ }\href
  {https://doi.org/10.1103/PhysRevLett.63.1253} {\bibfield  {journal} {\bibinfo
   {journal} {Phys. Rev. Lett.}\ }\textbf {\bibinfo {volume} {63}},\ \bibinfo
  {pages} {1253} (\bibinfo {year} {1989})}\BibitemShut {NoStop}%
\bibitem [{\citenamefont {Lyo}\ \emph {et~al.}(1989)\citenamefont {Lyo},
  \citenamefont {Kaxiras},\ and\ \citenamefont {Avouris}}]{LyoPRL1989}%
  \BibitemOpen
  \bibfield  {author} {\bibinfo {author} {\bibfnamefont {I.-W.}\ \bibnamefont
  {Lyo}}, \bibinfo {author} {\bibfnamefont {E.}~\bibnamefont {Kaxiras}},\ and\
  \bibinfo {author} {\bibfnamefont {P.}~\bibnamefont {Avouris}},\ }\bibfield
  {title} {\bibinfo {title} {Adsorption of boron on {Si(111)}: Its effect on
  surface electronic states and reconstruction},\ }\href
  {https://doi.org/10.1103/PhysRevLett.63.1261} {\bibfield  {journal} {\bibinfo
   {journal} {Phys. Rev. Lett.}\ }\textbf {\bibinfo {volume} {63}},\ \bibinfo
  {pages} {1261} (\bibinfo {year} {1989})}\BibitemShut {NoStop}%
\bibitem [{\citenamefont {Shi}\ \emph {et~al.}(2002)\citenamefont {Shi},
  \citenamefont {Radny},\ and\ \citenamefont {Smith}}]{ShiPRB2002}%
  \BibitemOpen
  \bibfield  {author} {\bibinfo {author} {\bibfnamefont {H.~Q.}\ \bibnamefont
  {Shi}}, \bibinfo {author} {\bibfnamefont {M.~W.}\ \bibnamefont {Radny}},\
  and\ \bibinfo {author} {\bibfnamefont {P.~V.}\ \bibnamefont {Smith}},\
  }\bibfield  {title} {\bibinfo {title} {Electronic structure of the
  {$\mathrm{Si}(111)\sqrt{3}\ifmmode\times\else\texttimes\fi{}\sqrt{3}R30\ifmmode^\circ\else\textdegree\fi{}\ensuremath{-}\mathrm{B}$}
  surface},\ }\href {https://doi.org/10.1103/PhysRevB.66.085329} {\bibfield
  {journal} {\bibinfo  {journal} {Phys. Rev. B}\ }\textbf {\bibinfo {volume}
  {66}},\ \bibinfo {pages} {085329} (\bibinfo {year} {2002})}\BibitemShut
  {NoStop}%
\bibitem [{\citenamefont {Andrade}\ \emph {et~al.}(2015)\citenamefont
  {Andrade}, \citenamefont {Miwa}, \citenamefont {Drevniok}, \citenamefont
  {Drage},\ and\ \citenamefont {McLean}}]{AndradeJPCM2015}%
  \BibitemOpen
  \bibfield  {author} {\bibinfo {author} {\bibfnamefont {D.~P.}\ \bibnamefont
  {Andrade}}, \bibinfo {author} {\bibfnamefont {R.~H.}\ \bibnamefont {Miwa}},
  \bibinfo {author} {\bibfnamefont {B.}~\bibnamefont {Drevniok}}, \bibinfo
  {author} {\bibfnamefont {P.}~\bibnamefont {Drage}},\ and\ \bibinfo {author}
  {\bibfnamefont {A.~B.}\ \bibnamefont {McLean}},\ }\bibfield  {title}
  {\bibinfo {title} {Surface and near surface defects in $\delta$-doped
  {Si(1{\hspace{0.167em}}1{\hspace{0.167em}}1)}},\ }\href
  {https://doi.org/10.1088/0953-8984/27/12/125001} {\bibfield  {journal}
  {\bibinfo  {journal} {Journal of Physics: Condensed Matter}\ }\textbf
  {\bibinfo {volume} {27}},\ \bibinfo {pages} {125001} (\bibinfo {year}
  {2015})}\BibitemShut {NoStop}%
\bibitem [{\citenamefont {Lee}\ \emph {et~al.}(2014)\citenamefont {Lee},
  \citenamefont {Ren}, \citenamefont {Jia},\ and\ \citenamefont
  {Cho}}]{LeePRB2014}%
  \BibitemOpen
  \bibfield  {author} {\bibinfo {author} {\bibfnamefont {J.-H.}\ \bibnamefont
  {Lee}}, \bibinfo {author} {\bibfnamefont {X.-Y.}\ \bibnamefont {Ren}},
  \bibinfo {author} {\bibfnamefont {Y.}~\bibnamefont {Jia}},\ and\ \bibinfo
  {author} {\bibfnamefont {J.-H.}\ \bibnamefont {Cho}},\ }\bibfield  {title}
  {\bibinfo {title} {Antiferromagnetic superexchange mediated by a resonant
  surface state in {Sn/Si(111)}},\ }\href
  {https://doi.org/10.1103/PhysRevB.90.125439} {\bibfield  {journal} {\bibinfo
  {journal} {Phys. Rev. B}\ }\textbf {\bibinfo {volume} {90}},\ \bibinfo
  {pages} {125439} (\bibinfo {year} {2014})}\BibitemShut {NoStop}%
\end{thebibliography}%

\clearpage
\begin{widetext}

\begin{center}
    \large {\bf Extended data figures and tables} \normalsize
\end{center}
\renewcommand{\figurename}{{\bf Extended data figure}}
\setcounter{figure}{0}

\begin{figure}[h]
    \centering
    \includegraphics[width=1\textwidth]{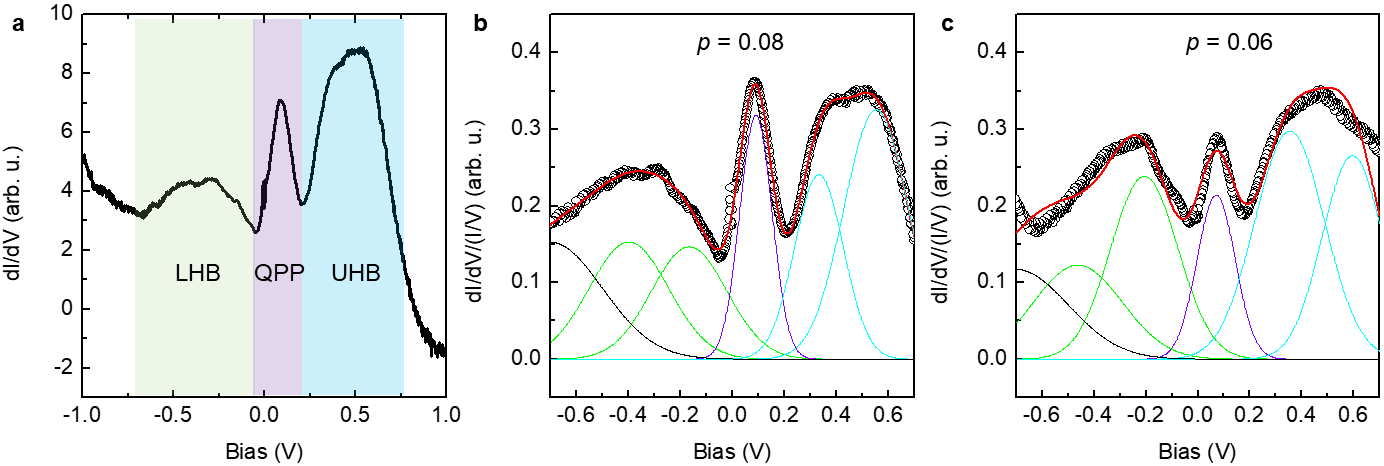}
    \caption{\small{\bf Spectral features of the ($\mathbf{\sqrt{3} \times\sqrt{3}}$)-Sn surface grown on two different Si wafers and estimates of the hole concentration}. {\bf a} $dI/dV$ spectrum of the $p = 0.08$ surface at 0.5 K, featuring the lower Hubbard band, quasi-particle peak, and upper Hubbard band (LHB/QPP/UHB).  {\bf b} $dI/dV/(I/V)$ spectrum obtained from the spectrum in panel {\bf a}, fitted with six Gaussian peaks. From the fitting, we find that the area under the QPP represents 16.1 \% of the total spectrum, excluding the peak on the far left which represents the contribution of the silicon valance band. This area fraction converts to a hole doping level of 8.05 \%, i,e, $p = 0.08$; See Ref.~\citenum{MingPRL2017} for more details. {\bf c} $dI/dV/(I/V)$ spectrum of the $(\sqrt{3}\times\sqrt{3})$-Sn surface (0.5 K), subject to the same fitting analysis as in {\bf b}. The area fraction of the QPP is 12.1 \%, which corresponds to hole doping level of 6.05 \% ($p = 0.06$). Gaussians are used only for the purpose of spectral area determination.}
    \label{fig:ED1_QPPband}
\end{figure}

\clearpage

\begin{figure}[h]
    \centering
    \includegraphics[width=1\textwidth]{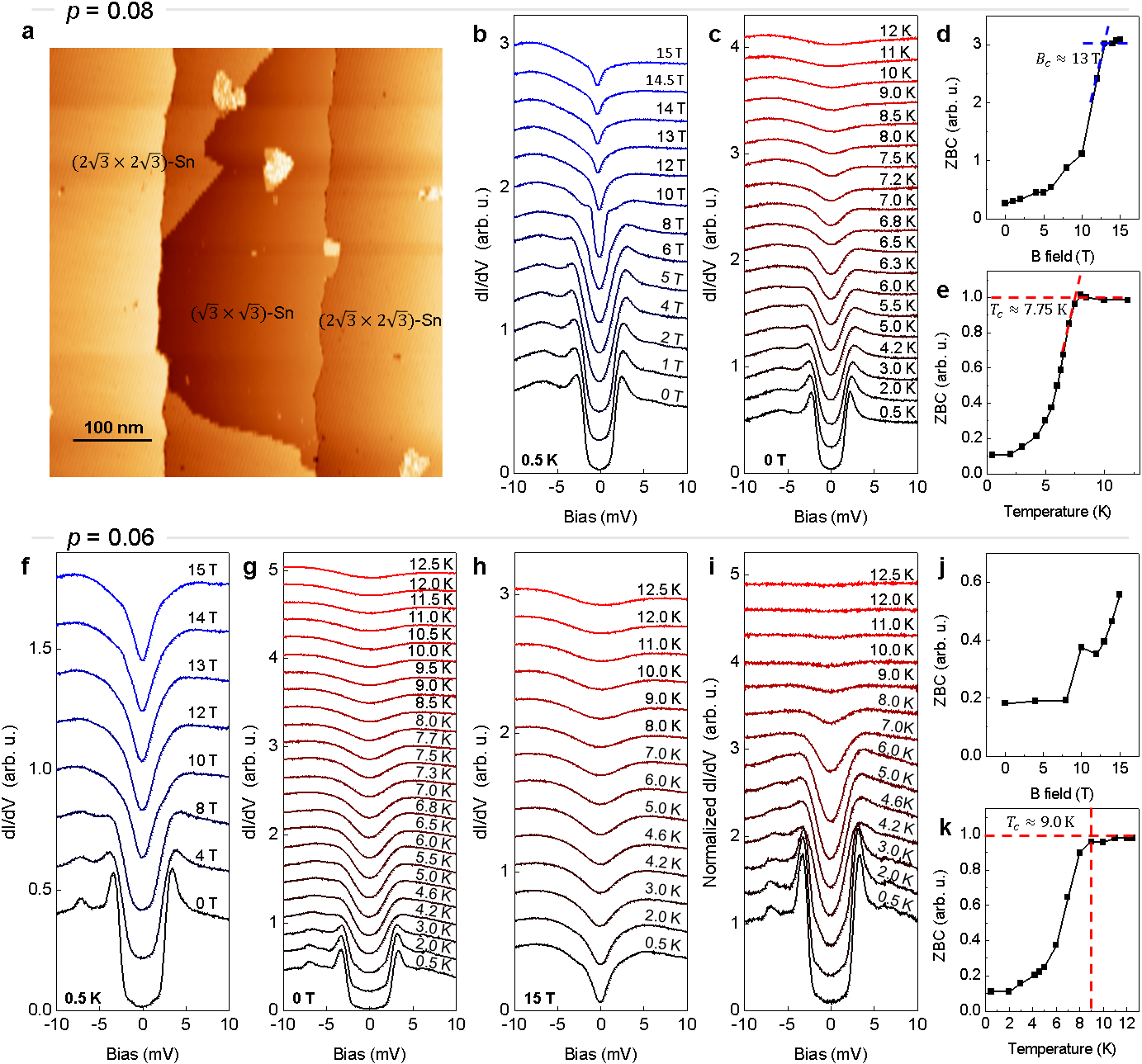}
    \caption{\small{\bf Tunneling spectroscopy of the superconducting state}. {\bf a} STM image ($V_s$ = -2 V, $I_t$ = 0.01 nA) of the $p = 0.08$ surface with neighboring (competing)  $(\sqrt{3}\times\sqrt{3})$-Sn and $(2\sqrt{3}\times2\sqrt{3})$-Sn domains. The former is superconducting while the latter is semiconducting. {\bf b}-{\bf e} STM tunneling spectra of the superconducting phase for $p = 0.08$. {\bf b}, Field dependent $dI/dV$ spectra measured at 0.5 K. {\bf c} Temperature dependent $dI/dV$ spectra measured in zero B-field. {\bf d} Zero bias conductance (ZBC) extracted from panel {\bf b}. The ZBC increases with the B field and saturates at $\sim 13$~T. {\bf e} ZBC extracted from normalized $dI/dV$ (most of the data are shown in Fig. \ref{fig:1}{\bf e}). The ZBC increases with the temperature and saturates around 7.8 K. {\bf f}-{\bf k} Tunneling spectra from the $p = 0.06$ surface. {\bf f} Field dependent $dI/dV$ spectra measured at 0.5 K. {\bf g} Temperature dependent $dI/dV$ spectra measured in zero B-field. {\bf h} Temperature dependent $dI/dV$ spectra measured in at 15 T. {\bf i}, $dI/dV$ spectra normalized by dividing the spectra in panel {\bf g} with the corresponding spectra in panel {\bf h} (same temperature), except for the 0.5 K and 2.0 K data in panel {\bf g}, for which we used  the 3.0 K data in panel {\bf h} so as to avoid division by the very small signal at zero bias. Note the persistence of the gap feature up to 9 K. {\bf j} ZBC extracted from panel {\bf f}. The ZBC increases with the B-field and does not saturate at 15~T. {\bf k} ZBC extracted from panel {\bf i}. The ZBC increases with temperature and saturates around 9 K.The normal state spectra in {\bf b} and {\bf c} exhibit minor suppression of the conductance near zero bias, which is due to the slow dissipation of the tunneling charge from the surface into the bulk, see Ref.~\citenum{MingPRB2018} for more details. Such effect becomes more significant for the $p = 0.06$ sample in panel {\bf f-g}.}
    \label{fig:superconducting_STS}
\end{figure}

\clearpage

\begin{figure}[h]
    \centering
    \includegraphics[width=1\textwidth]{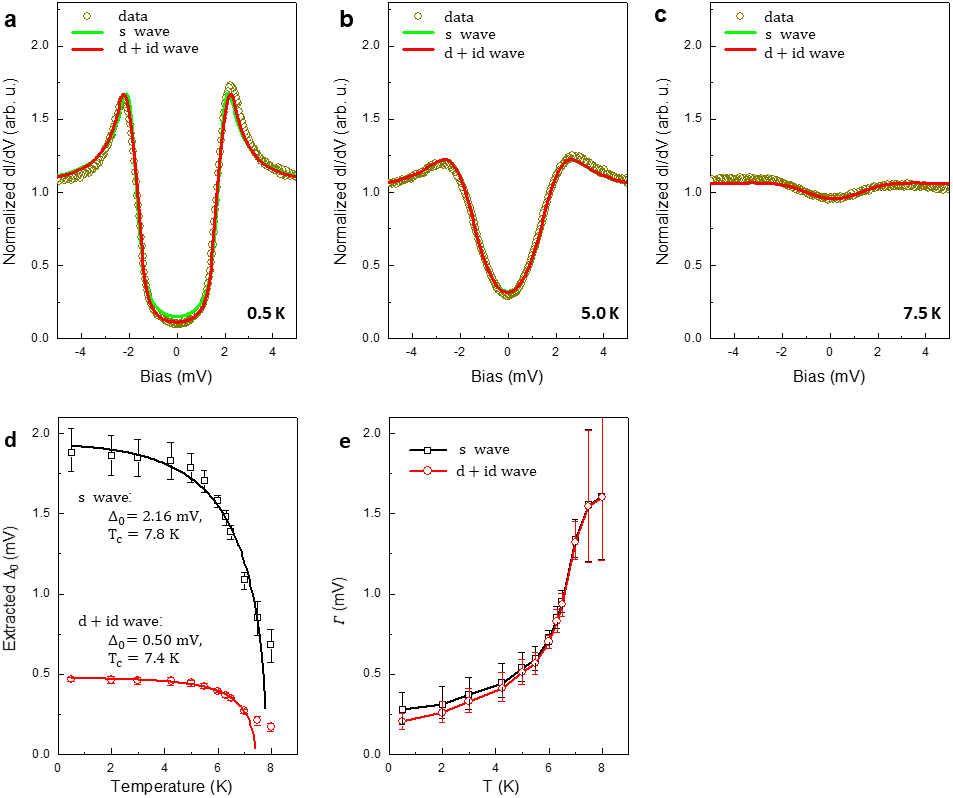}
    \caption{\small{\bf Fitting of the tunneling spectra}. 
    To fit the full $T$-dependence, we performed a Dynes-like fit of the $dI/dV$ spectra while adopting an angular-dependent gap function $\Delta(\theta)$ as parameterized in Ref.~\citenum{WuPRL2020}. (The results obtained using this approach are consistent with those obtained by fitting the full momentum-dependence Green's function in the superconducting state, see Fig.~\ref{fig:2}.) 
    {\bf a-c} Fitting results for the $p = 0.08$ system, assuming $s$-wave and $d_{x^2+y^2}+\mathrm{i}d_{xy}$ order parameters. The $s$-wave and $d_{x^2+y^2}+\mathrm{i}d_{xy}$-wave fits only reveal minor differences.   {\bf d} Extracted values of $\Delta_0$ as a function of temperature. {\bf e} The corresponding temperature dependence of the broadening parameter $\Gamma$. }
    \label{fig:fitting_2delta}
\end{figure}

\clearpage

\begin{figure}[h]
    \centering
    \includegraphics[width=1\textwidth]{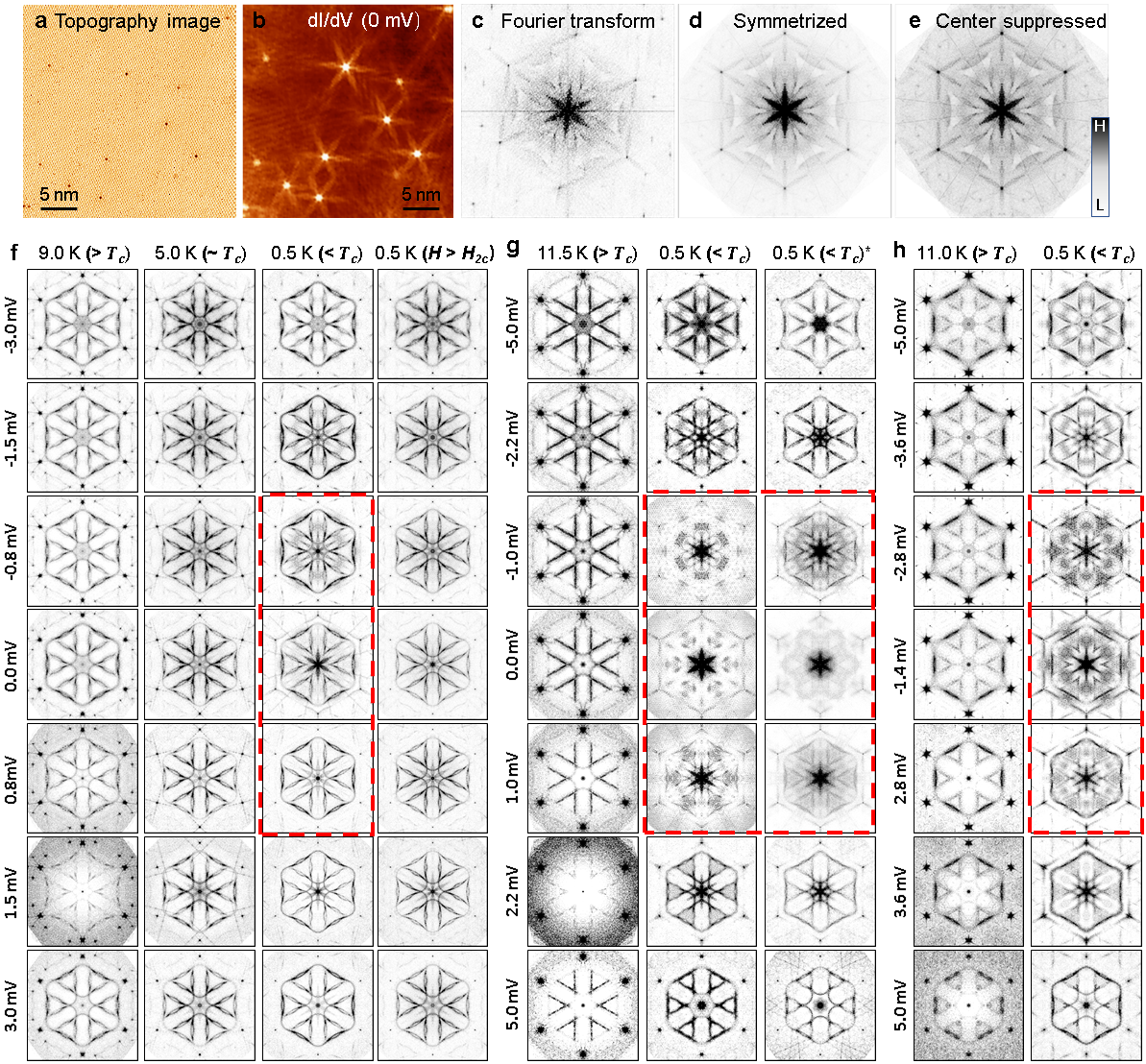}
    
    \caption{\small{\bf Experimental QPI results}. {\bf a}-{\bf e} QPI data and processing procedures. {\bf a} STM image ($V_s$ = 0.1 V, $I_t$ = 0.1 nA) of a ($\sqrt{3}\times\sqrt{3}$)-Sn surface ($p = 0.1$) with several surface defects appearing as dark spots. {\bf b} Corresponding $dI/dV$ image at $T = 0.5$ K. The bright star-like features are centered at the defect locations in panel {\bf a}. {\bf c} The power spectrum of panel {\bf b}, symmetrized and rotated in panel {\bf d}. The central region is subsequently suppressed to enhance the high frequency features, as shown in panel {\bf e} [see Ref.~\citenum{MingPRL2017} for more details]. {\bf f}-{\bf h} show 4, 3, and 2 sets of QPI results obtained from ($\sqrt{3}\times\sqrt{3}$)-Sn surfaces for $p = 0.1$, $0.08$, and $0.06$, respectively. Each column shows QPI images obtained in a fixed spatial region but with different biases, as indicated on the left. The measurement temperatures are labeled above each column, and data are shown for temperatures above and below $T_c$. The central flower leafs only appear when the sample is in superconducting state and when the measurement bias is within the superconducting gap (within $\pm 1.5$~mV, $\pm2.2$~mV, and $\pm3.6$~mV, in {\bf f}, {\bf g}, and {\bf h}, respectively). These QPI images are enclosed by the dashed red rectangles. Panel {\bf f} shows QPI results obtained at $T = 5~\mathrm{K}$ (slightly larger than $T_c = 4.7~\mathrm{K}$ for this sample), or at 0.5 K in an 8 T B-field ($H_{2c} = 3$~T). These data have a significantly reduced flower leaf feature, which could come from superconducting fluctuations. In panel {\bf g}, the ``0.5 K ($< T_c$)~$*$'' data are QPI results obtained from a sample with interstitial Sn adatoms, deposited at 120 K. The presence of interstitial Sn considerably enhances the flower-leaf features at the center of the Brillouin zone. }
    \label{fig:Experimental_QPI}
\end{figure}

\clearpage

\begin{figure}[h]
    \centering
    \includegraphics[width=0.7\textwidth]{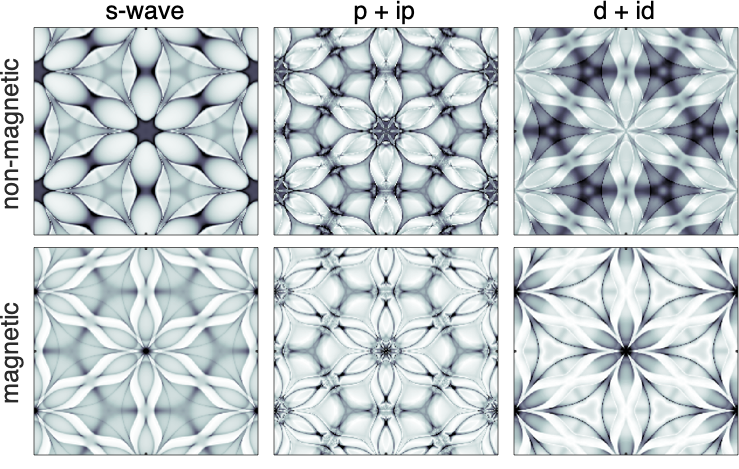}
    \caption{\small{\bf Simulated QPI spectra for different gap symmetries and scattering centers}. 
    The top and bottom rows show results for nonmagnetic ($\hat{V} = V_0\hat{\tau}_3$) and magnetic ($\hat{V} = V_0\hat{\tau}_0$) scatterers, respectively, with $V_0 = 100$ meV. Results are shown in the superconducting state assuming $s$-wave (left column), chiral $p+\mathrm{i}p$ (middle column), and $d+\mathrm{i}d$ (right column) order parameters. The spectra are calculated at a bias voltage of 1 meV. In each case, the magnitude of the gap $\Delta_0$  and smearing parameter $\delta$ are obtained from fits of the $dI/dV$ spectra shown in Fig.~\ref{fig:2}{\bf a} (see Methods). Note the absence of the central flower-leaf feature for non-magnetic scattering combined with the s-wave order parameter.
    }
    \label{fig:extended_QPI}
\end{figure}

\clearpage

\begin{figure}[h]
    \centering
    \includegraphics[width=1\textwidth]{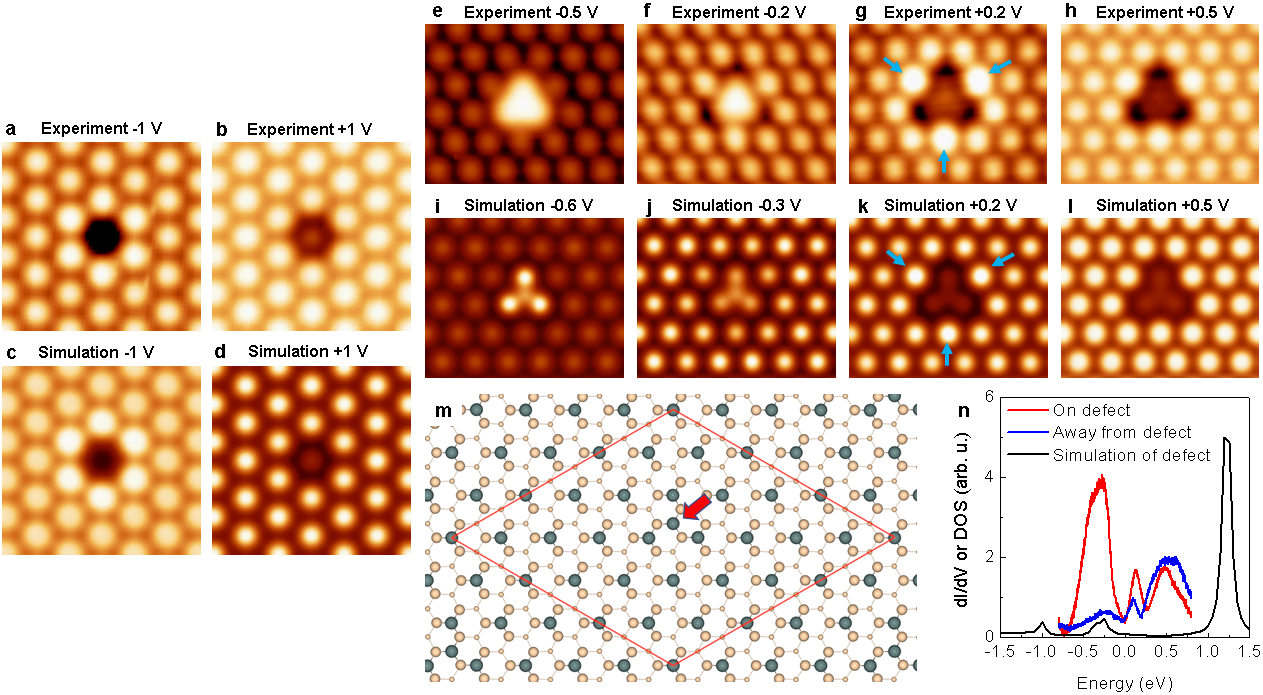}
    \caption{\small{\bf Experimental and simulated STM images of the substitutional Si and interstitial Sn adatom defects}. {\bf a-d} Experimental ($I_t = 0.1$ nA) and simulated STM images for the substitutional Si defect. The Si atom is invisible in filled state images, indicating that there are no occupied dangling bond states to tunneling from. The Si atom is visible in empty state images, but the atom appears to be dim due to its smaller covalent radius. {\bf e-h} Experimental STM images ($I_t = 0.1$ nA) of the interstitial Sn adatom defect center. Panel {\bf e} reveals three very bright adatoms that are part of the regular ($\sqrt{3}\times\sqrt{3}$)-Sn lattice. The interstitial Sn atom is located at the center of this cluster and cannot be imaged within the accessible bias range, as the adatom moves at higher biases. {\bf i-l} Simulated STM images. Note the slightly increased brightness of the Sn atoms indicated by the blue arrows in panel {\bf g}. This subtle effect is captured by the theory simulation in panel {\bf k}. {\bf m} ($9\times9$) supercell used in the DFT calculations for the interstitial Sn adatom defect. Sn adatom and Si substrate atoms are shown in green and gold, respectively. The interstitial Sn atom is placed near the center of the ($9\times9$) unit cell, as indicated by the red arrow. {\bf n} Experimental $dI/dV$ spectra recorded on top of the interstitial adatom (red) and far away from the interstitial location (blue). The latter reveals the characteristic LHB/QPP/UHB features (see Extended Data Fig.~\ref{fig:ED1_QPPband}). The strong peak at about -0.35 eV corresponds to the triangular adatom feature in panels {\bf a}, {\bf e}. It is captured by the DFT calculation (black line). The peak at $+1.25$ eV in the theoretical DOS mainly consists of the (empty) $5p$ orbitals of the interstitial adatom. Simulated images at this bias indeed visualize this atom (not shown), but it cannot be imaged at this tunneling bias.}
    \label{fig:Defect_and_Simulation}
\end{figure}

\clearpage

\begin{figure}[h]
\centering
\includegraphics[width=1\textwidth]{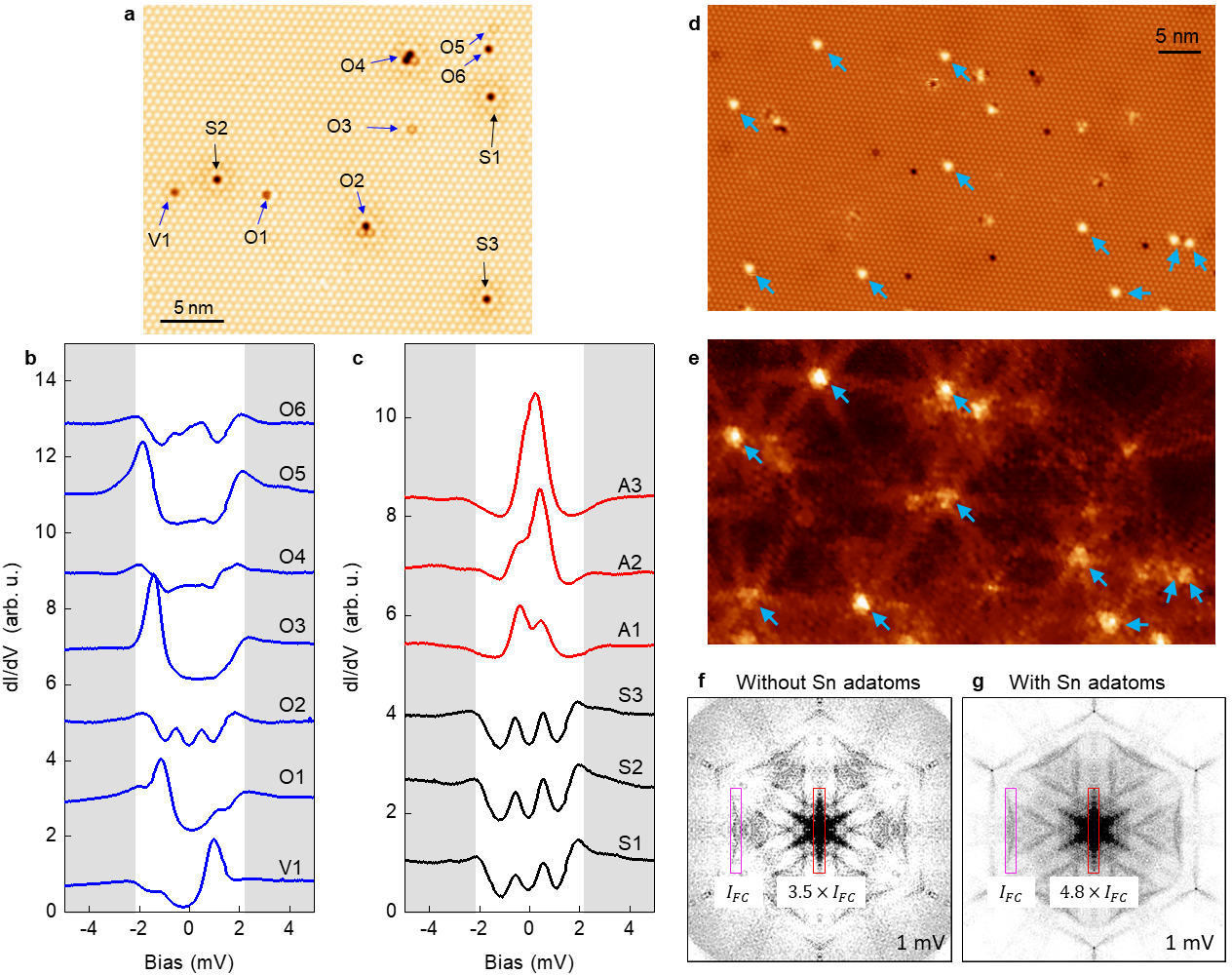}
\caption{\small{\bf Tunneling spectra (0.5 K) measured at defect locations on the superconducting $\mathbf{(\sqrt{3}\times\sqrt{3})}$-Sn surface ($\mathbf{p = 0.08}$) and corresponding QPI data}. {\bf a} STM image ($V_s$ = 0.02 V, $I_t$ = 0.1 nA) showing several intrinsic surface defects. S1, S2 and S3 are substitutional Si defects and V1 is a Sn vacancy, as judged by the appearance of a hole for a wide range of positive and negative tunneling biases (images at other biases are not shown). Defects with unknown structures are labelled On, with $n= 1-6$. {\bf b} Tunneling spectra of the V1 and On defects in panel {\bf a}.  {\bf c} Tunneling spectra of the interstitial Sn adatom defects A1, A2, and A3; see Supplementary Fig.~\ref{fig:Defect_and_Simulation}. All defects produce a pair of in-gap states. The substitutional Si defects (S1, S2, and S3) exhibit a well defined double-peak structure at $\pm{0.6}$ meV. Interstitial Sn adatoms (A1 and A2) also possess a double peak structure, but with smaller energy splitting ($\pm{0.2}$~meV). The A3 defect appears to have a single (unresolved) peak structure. {\bf d} STM image ($V_s = -0.5$~V, $I_t = 0.1$~nA) of a the $\mathbf{(\sqrt{3}\times\sqrt{3})}$-Sn surface with interstitial Sn defects (indicated by arrows) and other intrinsic defects, mostly substitutional Si or adatom vacancies. {\bf e}  Corresponding $dI/dV$ image obtained at a sample bias of $V_s=-0.4$~mV. The interstitial Sn defects produce the strongest scattering features in the real space conductance maps, as compared to those of other intrinsic defects. {\bf f},{\bf g} compares the QPI spectra from the same $\mathbf{(\sqrt{3}\times\sqrt{3})}$-Sn surface with and without adsorbed Sn defects. $I_{FC}$ represents averaged scattering intensities for the segment of the Fermi contour enclosed by the magenta rectangle, while the averaged intensity for the flower leaf features near the center of the Brillouin zone  (enclosed by the red rectangle) is represented in units of $I_{FC}$. The relative scattering intensity of the flower leaf feature is strongest for the surface with the interstitial Sn adatoms defects.
}
\label{fig:VariousDefects}
\end{figure}

\clearpage

\begin{figure}[h]
    \centering
    \includegraphics[width=0.7\textwidth]{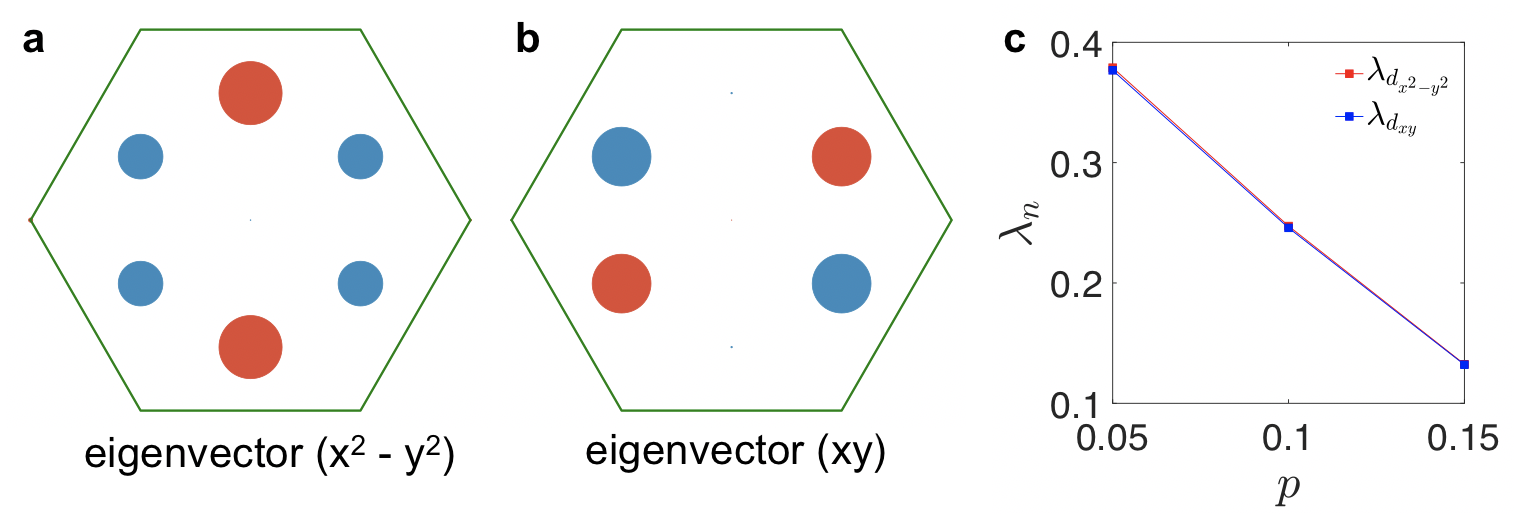}
    \caption{\small{\bf The leading (degenerate) eigenvectors of the Bethe-Salpeter equation for the triangular lattice Hubbard model}. {\bf a}, {\bf b} The momentum space structure of the leading eigenvectors, which show a pairing symmetry consistent with $d_{x^2-y^2}+\mathrm{i}d_{xy}$ pairing. The size  and color of the dots indicate the magnitude and sign of the eigenvector at the momentum points of the $3\times 3$ cluster. The 
    green hexagon shows the boundaries of the first Brillouin zone. {\bf c} The doping dependence of the leading eigenvalues at an inverse temperature of 
    $\beta = 8/t_1$. }
    \label{fig:dca_eigenvectors}
\end{figure}

\clearpage

\begin{figure}[h]
    \centering
    \includegraphics[width=1\textwidth]{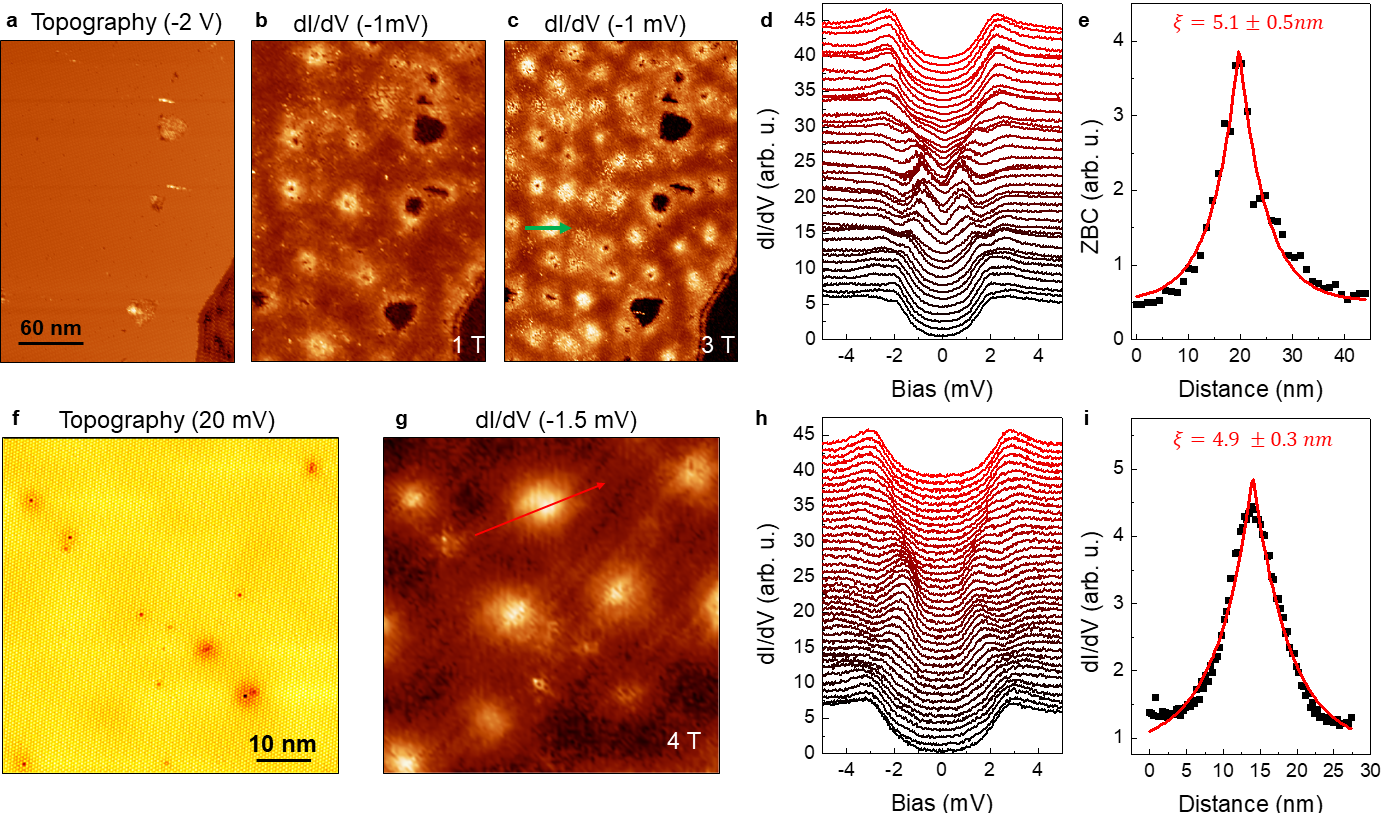}
    \caption{\small{\bf Magnetic vortices of the superconducting $\mathbf{(\sqrt{3}\times\sqrt{3})}$-Sn surface observed at 0.5 K}. {\bf a}-{\bf e} Magnetic vortices for the $p = 0.08$ sample. {\bf a} Topographic STM image. {\bf b}, {\bf c} $dI/dV$ maps obtained with a tunneling bias inside the superconducting gap with a B-field of 1 T and 3 T, respectively. {\bf d} $dI/dV$ spectra measured along the line crossing a magnetic vortex in panel {\bf c}. {\bf e} ZBC obtained from panel {\bf d} with an exponential fit for determining the superconducting  coherence length $\xi$. {\bf f}-{\bf i}  Magnetic vortices for the $p = 0.06$ sample. {\bf f} Topographic image. {\bf g} $dI/dV$ map obtained with a tunneling bias inside the superconducting gap with a B-field of 4 T. This sample exhibits a very low zero bias $dI/dV$ signal, presumably due to the very high series resistance of this lightly doped sample at 0.5 K. Therefore the vortex features are only resolved away from zero bias. {\bf h} $dI/dV$ spectra measured along the line crossing a magnetic vortex in panel {\bf g}. {\bf i} $dI/dV$ (- 1.5 mV) along the line indicated in panel {\bf g} with an exponential fit for determining the superconducting  coherence length.}
    \label{fig:vortex}
\end{figure}

\clearpage
\begin{figure}
    \centering
    \includegraphics[width=0.7\textwidth]{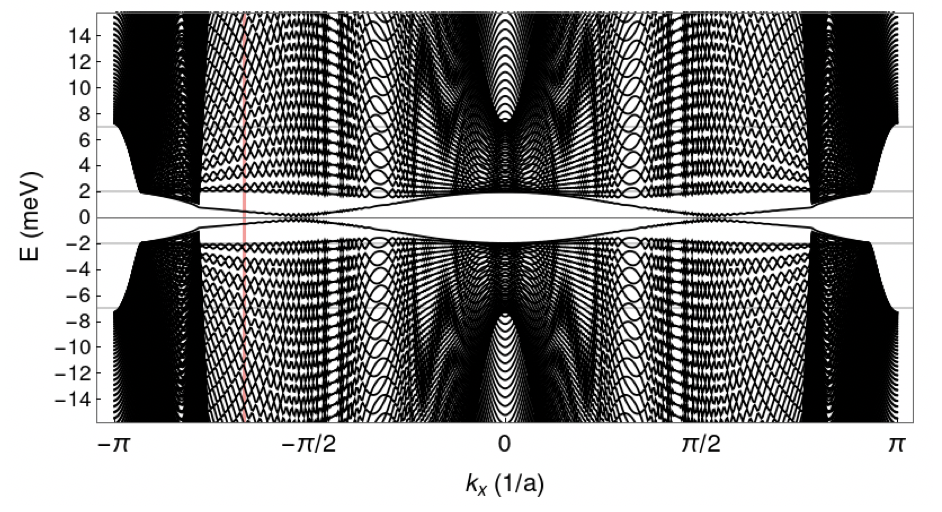}
    \caption{\small{\bf Electronic band dispersion for a chiral $d$-wave superconductor in the presence of edges.} Results were obtained by considering a mean-field $d_{x^2-y^2}+\mathrm{i}d_{xy}$ order parameter on a triangular lattice. The edge state spectrum is obtained by solving the tight-binding Hamiltonian on a cylinder, with open boundary conditions along the $y$-direction and periodic boundary conditions along the $x$-direction. Momentum $k_x$ remains a good quantum number and the resulting spectrum is shown here for a system of $400$ chains (labeled by $n$) stacked in the $y$ direction and a $k$-mesh of $400$ points. To clearly show the edge states here we computed the spectrum for $\Delta_0=1.5$ meV. As expected for a chiral $d$-wave superconductor with Chern number $C=\pm2$, two chiral linearly dispersing in-gap states exist on each edge. }
    \label{fig:edgemode}
\end{figure}

\clearpage 
\setcounter{figure}{0}
\setcounter{section}{0}
\setcounter{equation}{0}
\renewcommand{\figurename}{{\bf Supplementary Figure}}
\renewcommand{\thefigure}{{\bf \arabic{figure}}}%
\renewcommand{\theequation}{S\arabic{equation}}%

\begin{center}
    \large {\bf Supplementary Notes and Figures} \normalsize
\end{center}

\begin{center}
{\bf Supplementary Note 1: Modulation hole doping and dopant segregation}
\end{center}

Ming {\it et al.}~\cite{MingPRL2017} have shown that the formation of the ($\sqrt{3}\times\sqrt{3}$)-Sn surface is accompanied by a downward band bending near the surface of the boron-doped p-type Si substrate. This behavior means that the valence band holes in the subsurface region transfer to the dangling bond surface state. (Holes move up to lower their electronic energy; alternatively, the Sn adatoms give up some electrons to fill the hole states in bulk). This process is self-limiting, however, as a space charge layer dipole gradually builds up and prevents further charge transfer. (This band bending scenario is analogous to the formation of a depletion layer in semiconductor heterostructures.) 
The hole-doped surface state thus represents the total energy minimum. Carrier freeze-out in bulk Si at low temperature is irrelevant for our experiments as the relevant holes reside in a metallic surface state well above the valence band maximum (the quasi-particle band), which is formed during or after the growth stage (when cooling to room temperature). These holes are itinerant. 
 
The substrate cleaning procedure involves a high temperature annealing step that unavoidably leads to boron accumulation near the surface layers. Boron segregation ultimately produces a $\sqrt{3}\times\sqrt{3}$ surface {\it before depositing Sn}. Here, the boron atoms reside at the S5 site in the second layer  directly below the Si adatom, see e.g., Refs.~\citenum{HeadrickPRL1989} and \citenum{LyoPRL1989}. The dangling bond surface state of the Si adatom is empty; see Ref.~\citenum{LyoPRL1989}. Supplementary Figure~\ref{fig:SubstrateImages} provides images of the boron segregated substrates obtained before Sn deposition. Unless the boron is fully segregated, forming a $\sqrt{3}\times\sqrt{3}$ structure, the substrate surface can be quite inhomogeneous or it exhibits the native Si(111)$7\times 7$ reconstruction at the lowest doping level.  

Crucially, {\it after} depositing Sn at $600^\circ$C and subsequently cooling, the ($\sqrt{3}\times\sqrt{3}$)-Sn surface is very homogeneous (see Fig. 1 of the main text), both in topographic imaging and in $dI/dV$ imaging, which show only small amounts of disorder. Either the boron atoms diffuse deeper into the bulk during the formation of the ($\sqrt{3}\times\sqrt{3}$)-Sn surface at high temperature, or they form an ordered subsurface layer where the boron atoms occupy the S5 lattice locations. We cannot verify the first possibility directly, but the second possibility can be ruled out easily because the Sn-derived surface state band would be empty in this case (see Supplementary Note~3). Instead, we find it is almost half-filled. The latter can be inferred from the very close match between the theoretical and experimental $E({\bf k})$ dispersion relations, including the Fermi level crossings, as shown in Fig. 3k and 3l of Ref.~\citenum{MingPRL2017}. Our results are also fully consistent with a density functional theory (DFT) calculation where the boron atoms reside in deeper layers, and the surface state becomes hole-doped (see Supplementary Note~3).

\begin{figure}[h]
    \centering
    \includegraphics[width=\textwidth]{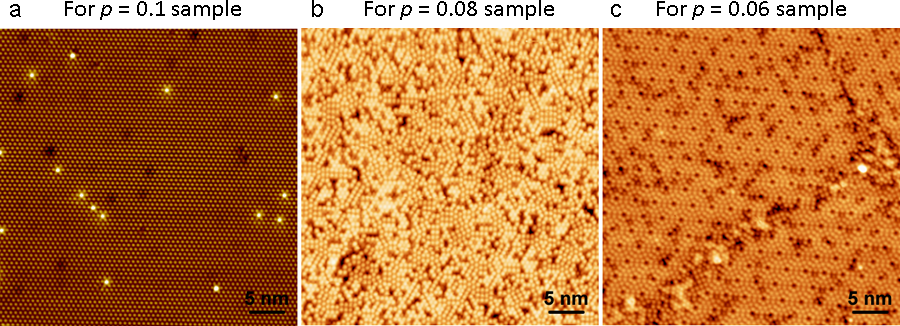}
    \caption{STM images of the boron segregated substrates obtained before Sn deposition. The hole concentration of the ($\sqrt{3}\times\sqrt{3}$)-Sn surface after the Sn deposition are labelled. {\bf a} The surface shows a near perfect $\sqrt{3}\times\sqrt{3}$ lattice indicating a saturated boron concentration (1/3 ML) near the surface. {\bf b} The surface is disordered at a lower boron concentration. {\bf c} The surface shows the native Si(111)($7\times 7$) reconstruction for the most lightly doped sample. }
    \label{fig:SubstrateImages}
\end{figure}

We reiterate that the hole-doped ($\sqrt{3}\times\sqrt{3}$)-Sn surface is structurally and electronically homogeneous (Fig. 1). Hole doping produces a single Fermi surface contour that is reproduced in our QPI image simulations based on the theoretical (DFT) band dispersion of the Sn-derived dangling-bond surface state. These observations demonstrate the efficacy of the modulation doping approach. Note that it would be inconceivable that a chemically disordered mix of Sn and B would produce such a result.

\begin{center}
    {\bf Supplementary Note 2: Additional fits to the STS spectra}
\end{center}

To rule out other possible pairing scenarios, we performed additional fits to the STS spectra assuming nematic $d_{x^2-y^2}$ and $d_{xy}$ order parameters, as well as an extended $p+\mathrm{i}p$ case. The latter case includes nearest- and next-nearest-neighbor pairing as parameterized in Ref.~\cite{ZhouPRL2008}. All three order parameters produce unsatisfactory fits of the STS spectra, as shown in Supplementary Figure~\ref{fig:STS_nematic}. The qualitative discrepancy here is caused by the fact that each of these order parameters produces a significant amount of anisotropy on the Fermi surface, with nodes for the $d$-wave cases.

\begin{figure}[h]
    \centering
    \includegraphics[width=0.7\textwidth]{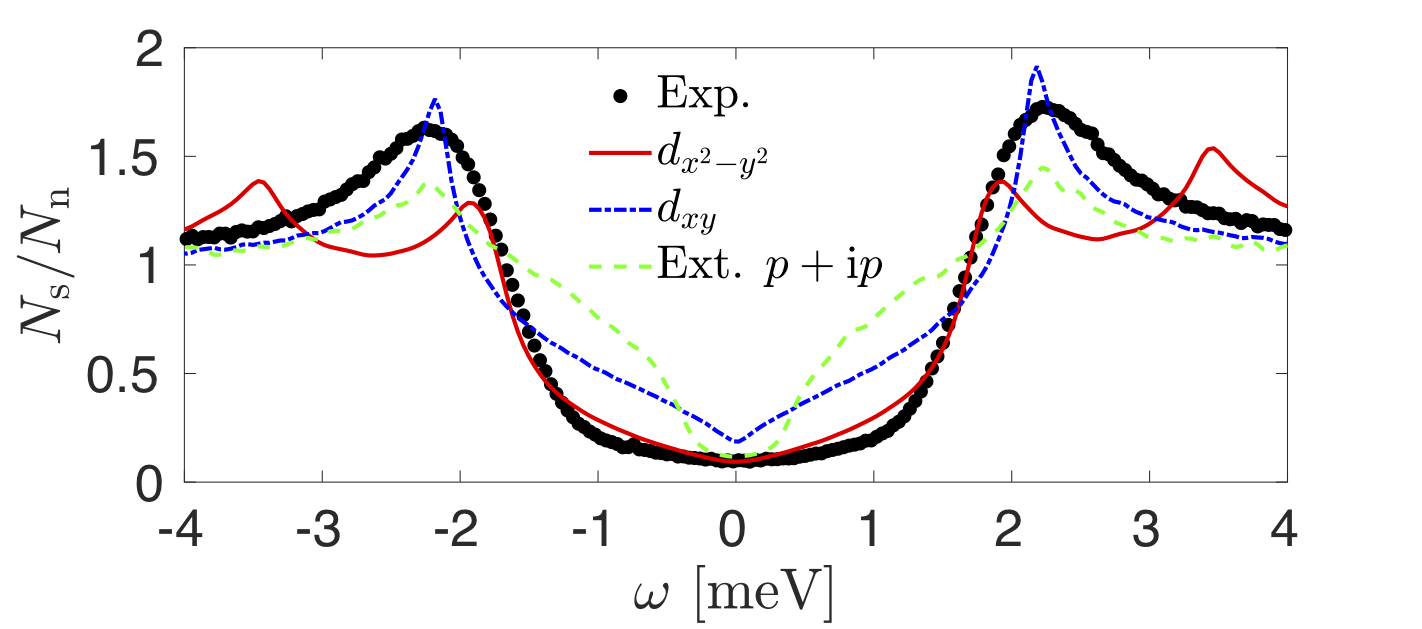}
    \caption{The normalized STS spectra fit with nematic $d$-wave order parameters and an extended $p+\mathrm{i}p$ order parameter with nearest- and next-nearest-neighbor pairing. }
    \label{fig:STS_nematic}
\end{figure}

We have also explicitly checked the scenario where the pairing interaction produces a momentum-dependent $s$-wave gap that resembles the chiral $d$-wave case. To do so, we introduced an $s$-wave gap defined as $\Delta_s({\bf k}) \equiv |\Delta_{d+\mathrm{i}d}({\bf k})|= \sqrt{\Delta^{\phantom*}_{d+\mathrm{i}d}({\bf k})\Delta_{d+\mathrm{i}d}^*({\bf k})}$. 
This substitution produces a real valued gap $\Delta_s({\bf k})$ whose momentum dependence $\Delta^2_s({\bf k})$ perfectly mimics the anisotropy of the chiral $d$-wave case $|\Delta_{d+\mathrm{i}d}({\bf k})|^2$ but lacks the phase information. Defined in this way, $\Delta_s({\bf k})$ is a six-fold symmetric $s$-wave order parameter that produces an STS fit indistinguishable from our chiral $d$-wave case. This case does not reproduce the flower-like features observed in the QPI spectra. These results make it clear that the data are inconsistent with an anisotropic $s$-wave, chiral $p$, and nematic $d$-wave order parameters. \\

\begin{center}
{\bf Supplementary Note 3: DFT calculations for the B-doped surface}
\end{center}

DFT calculations for the boron saturated ($\sqrt{3}\times\sqrt{3}$) substrate surface (without Sn) can be found in Refs.~\citenum{ShiPRB2002} and \citenum{AndradeJPCM2015}. The surface is gapped and the only state in the gap is the adatom dangling state that is located well above the Fermi level near the conduction band continuum. 

In this work, we performed new plane-wave DFT calculations for a 10-layer Si slab terminated with the ($\sqrt{3}\times\sqrt{3}$)-Sn structure on one side and saturated with hydrogen on the other. We considered the following scenarios for the ($\sqrt{3}\times\sqrt{3}$)-Sn surface: 
\begin{enumerate}[label=\alph*)]
    \itemsep0em 
    \item {no boron doping;}
    \item {one boron atom per $\sqrt{3}\times\sqrt{3}$ unit cell, where the boron is residing directly below the Sn adatoms; }
    \item {one boron atom per $\sqrt{3}\times\sqrt{3}$ unit cell, where the boron atom is located in the fifth Si layer (corresponding to an extremely high bulk concentration of about $10^{21}$ boron atoms per cm$^3$); and}
    \item{one boron atom per $2\sqrt{3}\times 2\sqrt{3}$ unit cell, again with the boron atom residing in the fifth layer.}
\end{enumerate}
(In the $2\sqrt{3}\times 2\sqrt{3}$ calculations, we did not consider the scenario with one boron atom located at one of the four S5 lattice locations in the unit cell as this would produce inequivalent lattice sites within the $2\sqrt{3}\times 2\sqrt{3}$ unit cell, whereas the lattice seen by STM is a perfect ($\sqrt{3}\times\sqrt{3}$). Results are shown in Supplementary Figure~\ref{fig:new_DFT}. 

\begin{figure}[b]
    \centering
    \includegraphics[width=\textwidth]{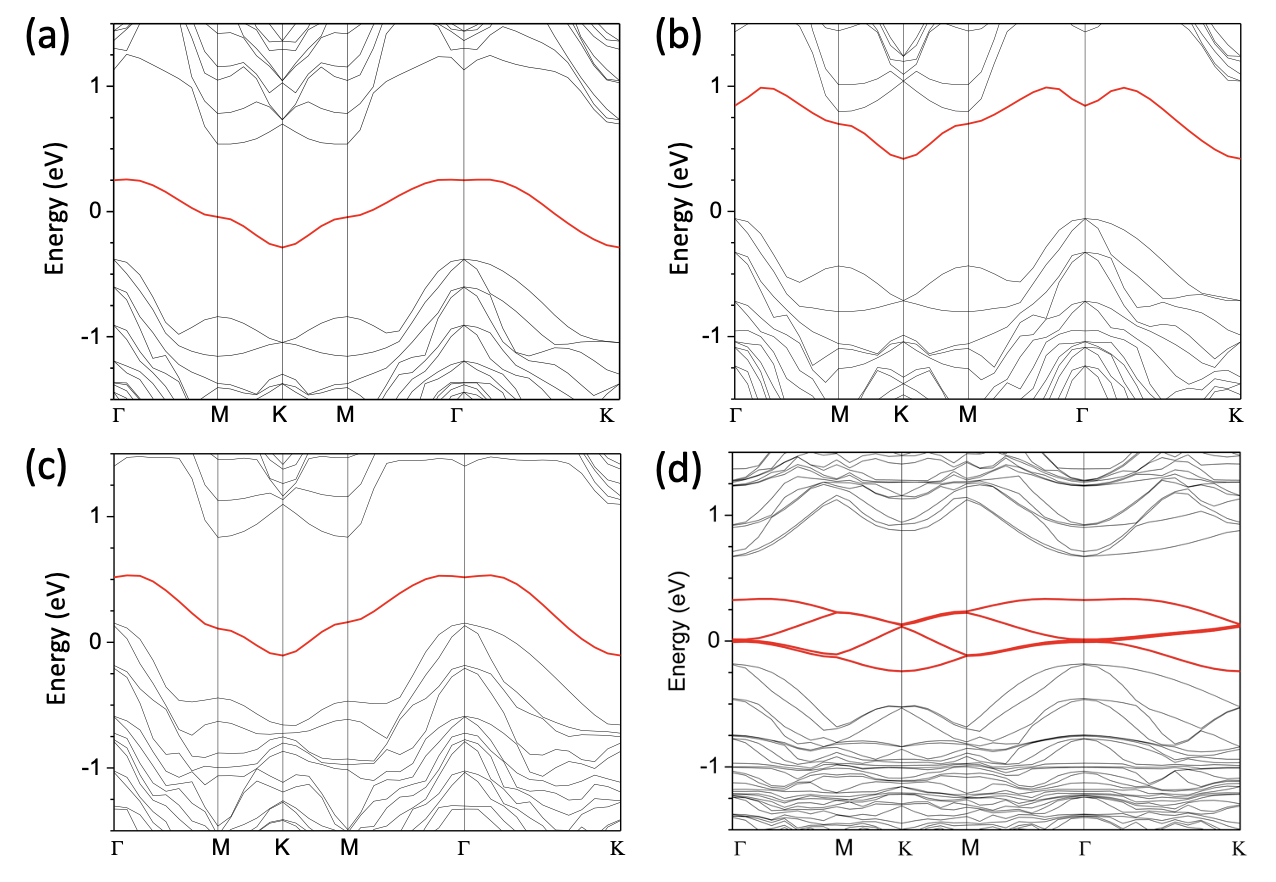}
    \caption{DFT calculations for the noninteracting band-structure of  the Sn/Si(111) system. Results are shown for (a) no boron; (b) one boron atom per $\sqrt{3}\times\sqrt{3}$ unit cell, where the boron is residing directly below the Sn adatoms; (c) one boron atom per $\sqrt{3}\times\sqrt{3}$ unit cell where the boron atom is located in the fifth Si layer; and (d) one boron atom per $2\sqrt{3}\times 2\sqrt{3}$ unit cell, again with the boron atom residing in the fifth layer. }
    \label{fig:new_DFT}
\end{figure}

Scenario a) (Supplementary Figure~\ref{fig:new_DFT}a),  produces a single Sn-derived dangling bond surface state inside the bulk band gap of Si. The band is half-filled (as expected with one electron per dangling bond site) and is subject to the Mott transition resulting in the upper and lower Hubbard band seen in experiment; see e.g. Refs.~\citenum{MingPRL2017} and \citenum{LiNatComm2013}. 

In scenario b) (Supplementary Figure~\ref{fig:new_DFT}b), the Sn-derived dangling bond surface state is located entirely above the Fermi level. Here, the dangling bond electrons are fully compensated by the holes in the subsurface layer, and the surface is semiconducting. To put it differently, each boron atom at the S5 lattice location accepts the dangling bond charge of the Sn atom sitting right above and becomes a negative acceptor ion. This scenario is inconsistent with the observation of the remnant lower and upper Hubbard bands in the STS spectra and the observation of a dispersive QPP state in the QPI spectra in the doped system. 

Scenario c) (Supplementary Figure~\ref{fig:new_DFT}c) reveals the same band, but this time it is crossing the Fermi level. The Fermi level is about 0.1 eV below the van Hove singularity at the $M$-point. This situation represents a heavily overdoped case, an artifact of the small supercell used in the calculation. Experimentally, the Fermi level is located 7 meV {\it above} the van Hove singularity ~\cite{MingPRL2017}. The calculation furthermore places the Fermi level below the top of the bulk Si valence band, while in the experiment, the Fermi level is located 0.5 eV above the valence band maximum; see supplementary information in Ref.~\citenum{MingPRL2017}. 

Finally, scenario d) (Supplementary Figure~\ref{fig:new_DFT}d) is approximately consistent with experimental observations. Because the unit cell was quadrupled in this calculation, the doping level is reduced by a factor of four, and there are now four Sn bands inside the Si band gap instead of one. These bands originate from the same Sn-derived dangling bond surface state, backfolded about the $2\sqrt{3}\times 2\sqrt{3}$ Brillouin zone. (In experiment, we only see one band as the translational symmetry is $\sqrt{3}\times \sqrt{3}$.) Importantly, the Fermi level is about 0.18 eV above the valence band maximum, and the band filling is $3/8$. This concentration corresponds to a hole doping level of $1/8$ or $x=0.125$, still exceeding the experimentally determined doping level by about a factor of two. Due to the Mott correlations, the surface state splits into an upper and lower Hubbard band and, because the surface state is no longer half-filled, a quasi particle peak appears in between; see Ref.~\citenum{MingPRL2017}. A similar calculation for a $3\sqrt{3}\times 3\sqrt{3}$ unit cell (not shown) produces a hole concentration of $1/18$ or $x=0.056$, which is close to that of the $p=0.06$ sample. Here the Fermi level is about 0.30 eV above the valence band maximum. Calculations for lower doping levels would require even larger super cells, but the trend is clear: The holes are all located in the Sn dangling bonds, except at extremely high doping levels where the Fermi level crosses the valence band maximum, creating holes at the top of the valence band near the $\Gamma$-point. Experimentally, however, the valence band maximum is located 0.5 eV below the Fermi level, meaning there are no bulk valence band holes.

The important conclusion here is that these different doping scenarios affect the location of the Sn-derived surface state and chemical potential inside the gap. The only band crossing the Fermi level is the Sn-derived dangling bond surface state at the experimentally realizable doping levels. While there are many other possible distributions of the boron atoms, it is critically important to recognize that the surface is structurally and electronically homogeneous (see Fig. 1, main text). The boron dopants are likely located in deeper layers, not at the S5 lattice locations of the second layer (scenario b).

The DFT calculations also rule out the possibility of a boron impurity band crossing the Fermi level. The fact that the DFT calculations do not produce any B-derived bands in the Si band gap, indicates that the B impurity states are not in the band gap. There is no evidence of coherent hopping between B sites, which would also be very unlikely for a disordered boron distribution, as the potential would no longer be periodic.

\begin{center}
{\bf Supplementary Note 4: Nonmagnetic impurities in a magnetically ordered background}
\end{center}

One might argue that because the {\it undoped} Sn/Si(111) system has a $2\times 1$ striped antiferromagnetic order \cite{LiNatComm2013, LeePRB2014}, magnetic order might persist upon hole doping, in which case a nonmagnetic defect could be viewed as a magnetic perturbation. For this scenario to be true, the antiferromagnetism would have to coexist with superconductivity within a superconducting domain. There are no known examples of such coexistence in $s$-wave superconductors. Hence, this  would seem incompatible with a conventional $s$-wave order parameter. Instead, it would suggest an unconventional pairing symmetry. A chiral state is then naturally expected for the triangular lattice geometry. 

Finally, we note that the size and shape of the Fermi surface inferred from the QPI spectra for the doped system is consistent with DFT calculations for a paramagnetic Sn lattice; it does not have any indications of zone-folding that must accompany any antiferromagetic ordering.

\end{widetext}
\end{document}